\documentstyle[amsfonts,epsfig,preprint,aps]{revtex}

\begin{document}
\draft
\preprint{ }

\title{A thermodynamically self-consistent theory for the Blume-Capel model.}
\author{S. Grollau, E. Kierlik, M. L. Rosinberg, and G.Tarjus}
\address{Laboratoire de Physique Th{\'e}orique des Liquides \cite{AAAuth},
Universit{\'e} Pierre et Marie Curie,\\ 
4 Place Jussieu, 75252 Paris Cedex 05, France}
\date{\today}
\maketitle

\begin{abstract}
We  use a self-consistent Ornstein-Zernike  approximation to study the
Blume-Capel ferromagnet on three-dimensional lattices. The correlation
functions and the thermodynamics are obtained from the solution of two
coupled  partial   differential equations.   The   theory  provides  a
comprehensive and accurate  description of  the  phase diagram in  all
regions,  including the wing boundaries  in non-zero magnetic field. In
particular, the coordinates of the tricritical point  are in very good
agreement with the best estimates from  simulation or series
expansion. Numerical and analytical analysis strongly suggest that the
theory predicts a universal Ising-like critical behavior along the
$\lambda$-line and the wing critical lines, and a tricritical behavior
governed by mean-field exponents.
\end{abstract}
\pacs{{\bf Key words}: Blume-Capel model, Ornstein-Zernike
approximation, tricriticality.}

\newpage

\def\be{\begin{equation}}
\def\ee{\end{equation}}
\def\bea{\begin{eqnarray}}
\def\eea{\end{eqnarray}}
\def\bit{\begin{itemize}}
\def\eit{\end{itemize}}

\section{Introduction}

The self-consistent  Ornstein-Zernike  approximation (SCOZA)  has been
introduced some time  ago by Hoye  and Stell \cite{HS1977} as a method
for obtaining thermodynamic  and structural properties of simple fluid
and lattice-gas systems.   Like the mean-spherical approximation, this
approach  is  based  on  the assumption that   the  direct correlation
function   $C({\bf  r})$,   which  is  related   to   the two-particle
distribution function    $G({\bf r})$  via  the Ornstein-Zernike  (OZ)
equation, is proportional to the pair  potential outside the hard-core
region  (or, for a  lattice-gas, for ${\bf r}  \neq  {\bf 0}$).  But the
dependence of the proportionality  constant on density and temperature
is determined in such a way that the same free energy is obtained from
fluctuation theory  - the so-called  compressibility or susceptibility
route  - and from integration of  the internal energy  with respect to
the  inverse temperature.   For  the lattice-gas with nearest-neighbor
attractive   interactions  (or equivalently,  for   the  ferromagnetic
spin-$\frac{1}{2}$ Ising  model), this  thermodynamic self-consistency
is embodied in a  partial differential equation whose solution,  along
with the requirement of single site occupancy, fixes $C({\bf r})$ (and
thus $G({\bf r})$ ) uniquely.  Because of numerical difficulties, this
equation was only solved recently\cite{DS1996}, showing that the SCOZA
provides an accurate description of  the properties  of the 3-d  Ising
model over most  of the phase diagram.   The predicted values of $T_c$
for the   various cubic lattices  are within  $0.2  \%$  of their best
estimates, the effective critical  exponents are faithfull to the true
behavior  above $T_c$  except in  a  very narrow  neighborhood of  the
critical    point,  and  the    zero-field  magnetization is described
asymptotically           by     the      nonclassical         exponent
$\beta=0.35$\cite{HPS2000}. The SCOZA   has  been also extended  to  {\it
n}-component and  continuous spins\cite{HS1997a}, and accurate results
have been obtained for   the hard-core Yukawa fluid\cite{PSW1998}  and
for       several  spin  systems    in     the  presence of   quenched
disorder\cite{KRT1997}.

The purpose of this paper  is to apply  the same type of approximation
to the Blume-Capel model \cite{BC1966},  a special case of the  spin-1
Blume-Emery-Griffiths (BEG)  model\cite{BEG1971}, which  represents  a
variety of interesting  physical systems, in particular $^3He$ -$^4He$
mixtures.  This model has played an important  role in the development
of the theory of tricritical phenomena
\cite{RW1972,LS1984} and has  been very  actively studied over
the years, following the     original  mean field treatment  of    BEG
\cite{BEG1971}. Various methods like series expansions \cite{FG1972},
renormalization-group    calculations\cite{BW1976},  and   Monte-Carlo
simulations\cite{JL1980} have been   used to describe the  first-  and
second order regions,     the tricritical region,  and the   crossover
between   them.    A  study  of   the coexistence    curve  using  the
mean-spherical    approximation    has   also        been     proposed
recently\cite{HS1997b}.  There is no analytical theory, however, which
is able  to provide a   comprehensive and accurate description  of the
phase  diagram  in all regions  (including  the ``wing'' boundaries in
non-zero magnetic field).  As we shall see in the following, the SCOZA
reaches this goal  quite  successfully whithout  requiring prohibitive
computational effort.    In  particular,   the   coordinates of    the
tricritical point (TCP)  for the various  cubic lattices are predicted
with very good accuracy  and the universal  asymptotic tricritical behavior
is well described.  This suggests that  the SCOZA is a reliable theory
for exploring three-dimensional  systems which exhibit  first-order as
well as continuous transitions.

The paper is organized as follows: in section 2 we describe the theory
and  derive   the  partial  differential  equations  that   encode the
thermodynamics of the  model, in section 3 we  present our results for
the   phase diagram,   and  in  section  4   we discuss  the universal
properties   in  the  second-order  and    tricritical regions.    Our
conclusions are  drawn in section 5.   The extension  of the theory to
the full  spin-$1$ Hamiltonian is   presented  in Appendix  A  and
details on the scaling behavior near the TCP are reported in Appendix B.

\section{Theory}
 
The Blume-Capel (BC) model\cite{BC1966} is defined by the Hamiltonian
\be
{\cal H}_{BC} = -J \sum_{<ij>} S_{i}S_{j}  +\Delta \sum_i S_i^2 -h \sum_i S_i 
\ee
where $S_i= 0,\pm  1$   is the  spin variable  at   each  site i  of   a
d-dimensional  lattice   and   the    first    term sums     over  all
nearest-neighbor (n.n.)    pairs.   This is a    special case  of  the
Blume-Emery-Griffiths   (BEG)     Hamiltonian\cite{BEG1971},    ${\cal
H}_{BEG}={\cal    H}_{BC}-  K \sum_{<ij>}S_{i}^2S_{j}^2$,   which   is a
microscopic  model  for  $^3He-^4He$  mixtures.  $S_i=0$ represents  a
$^3He$ atom at site $i$ and  $S_i=\pm1$ a $^4He$  atom, with the sign in
the latter case describing the superfluid degree  of freedom.  In this
interpretation, the coupling   constant   $J>0$ is a  potential   that
promotes  superfluidity,   the    crystal-field  $\Delta$   reflects   the
chemical-potential difference between the isotopes, and $K$ represents
the difference in the van  der Waals interactions between the isotopes
(the actual  value of $K/J$  is  small, so that   setting $K=0$  is  a
sensible approximation).   The magnetization  $m=<S_i>$ identifies  to
the superfluid order parameter and $x=1-<S_i^2>$ represents the $^3He$
concentration.   As is well  known,  the phase  diagram of $^3He-^4He$
mixtures  presents a line  of second-order  transitions (the so-called
$\lambda$-line) at high temperatures  and high $^4He$ concentrations and  a
coexistence   region associated with a  first-order  transition at low
temperatures.    The   BC   Hamiltonian    may    also    describe   a
spin-$\frac{1}{2}$ Ising model with a  fractional concentration $x$ of
non-magnetic impurities in thermal  equilibrium  with the spin  system
(annealed  dilution).    $\Delta$ is   then  interpreted  as the  chemical
potential   that controls the impurity   concentration (the  case of {
quenched dilution has been studied in Ref. \cite{KRT1997}).

Our theory  for   the  Blume-Capel model   is  based  on  an  Ornstein-Zernike
approximation for the  direct correlation function $C_{ij}$  which is
 the    inverse  of the   connected    pair  correlation function
$G_{ij}=<S_iS_j>-<S_i><S_j>$, i.e.,

\be \sum_kG_{ik}C_{kj}=\delta_{ij} 
\ee
where $\delta_{ij}$ is the Kronecker symbol. This OZ  equation may be
considered as the definition of $C_{ij}$. It is also a consequence 
of the  Legendre transform 
\be
{\cal G}={\cal F}+\sum_i h_im_i 
\ee
that  defines the Gibbs  free energy ${\cal  G}$  from the free energy
${\cal F}=-k_BT\ln Tr \exp[ -  {\cal H}_{BC}/(k_BT)]$.  In Eq.  (3), a
site-dependent magnetic  field    $h_i$   has been   introduced    for
convenience   and $<S_i>=- \partial{\cal  F}/ \partial    h_i =m_i$  is the  local
magnetization.  The second   functional derivatives of ${\cal  F}$ and
${\cal G}$ with  respect to the local  fields and local magnetizations
generate $G_{ij}$ and $C_{ij}$, respectively,
\be
G_{ij}=- \frac{\partial ^2 \tilde {\cal F}} {\partial{\tilde h_i} \partial {\tilde h_j}}
\ee
\be
 C_{ij}=  \frac{\partial ^2 \tilde  {\cal G}}{\partial m_i  \partial m_j} 
\ee
where ${\tilde {\cal F}}={\beta \cal F },{\tilde {\cal G}}={\beta \cal G }$,
and $\tilde h_i=\beta h_i$ ($\beta=(k_BT)^{-1}$ is the inverse temperature).
For a uniform  magnetic field $h_i=h$,  the system is  translationally
invariant  and  the correlation  functions only  depend  on the vector
${\bf r}$ that connects the two sites. The OZ equation then simplifies
to

\be
 \hat C({\bf k}) \hat G({\bf  k})=1  
\ee
in Fourier space.

In contrast   with $G({\bf r})$,  the  direct  correlation function is
expected to remain a short-ranged function even in the critical region
(specifically, $\hat C({\bf k}={\bf 0})<  +\infty$).  In  the following, we  shall
assume  that  $C({\bf r})$   has always the  same   range as  the pair
potential.   This OZ  ansatz is  the {\it  only} approximation  in our
theory.  Since the exchange interaction in the Blume-Capel Hamiltonian
is limited to nearest-neighbor sites, this amounts to setting

\be  
C({\bf r})= c_0(\tilde J,\tilde \Delta,m) \delta_{{\bf r},0} + c_1(\tilde J,
\tilde \Delta,m) \delta_{{\bf r},{\bf e}}
\ee
or in Fourier space,
\be  \hat C({\bf k})=c_0(\tilde J,\tilde \Delta,m)[1-z(\tilde J,\tilde
\Delta,m)\hat \lambda({\bf  k})]
\ee 
where ${\bf e}$  is a  vector from the   origin to one of its  nearest
neighbors, $\hat \lambda({\bf  k})=\frac{1}{c} \sum_{{\bf  e}} e^{i{\bf k.e}}$ is
the  characteristic function of  the lattice  ($c$ is the coordination
number), and $z=-\frac{c_1}{c_0}c$.    $c_0$ and $c_1$   (or,  equivalently,  $c_0$ and $z$)  are
functions of $\tilde  J=\beta J,\tilde \Delta=\beta \Delta$  and $m$, which will  be
obtained  below from the  solution of  the  SCOZA partial differential
equations  (in  the simpler  mean-spherical  approximation  studied in
ref.\cite{HS1997b},  one  has just   $c_1=\tilde  J$).   It  is  worth
noticing that  the range  of   $C(\bf{r})$ is exactly  limited  to n.n.
separation in one dimension.  This can be  easily checked by using the
transfer matrix  method  (see e.g.  ref.\cite{KF1975}) to  calculate $
G({\bf r})$  and  then inverting the OZ  equation   to get the  direct
correlation function (this result holds  for  the most general  spin-1
Hamiltonian  with  n.n. pair   interactions  ${\cal  H}={\cal H}_{BC}-
K\sum_{<ij>}S_{i}^2S_{j}^2-L\sum_{<ij>}S_{i}S_{j}(S_i+S_j)$ which  is used
as a model  for ternary mixtures\cite{MB1974}). $C(\bf{r})$ has
the same spatial structure in  the limit of infinite dimension  (where
the  mean-field approximation becomes exact), and we  expect that Eq.  (8)
is  a reasonable assumption  for  $d=3$.  A  major consequence is that
$G({\bf r})$ is given in any dimension by
\be   
G({\bf r})=\frac {1} {c_0}  P({\bf  r},z) 
\ee
where 
\be
 P({\bf r},z)= \frac {1} {{(2\pi)}^d} \int_{-\pi}^{\pi}d{\bf k}\frac{e^{-i{\bf k}.{\bf r}}}{1-z\hat{\lambda}({\bf k})} 
\ee
is  the lattice  Green's    function (for  instance, $   \hat  \lambda({\bf
k})=\frac{1}{3}(\cos  k_x+\cos  k_y+\cos k_z)$  for  the  simple cubic
lattice).  The variable $z$  (with  $0\leq  z\leq 1$)  is related  to  the
second-moment correlation length $\xi$ defined by $\hat G({\bf k})\sim
\hat G({\bf 0})(1+\xi^2k^2)$, $k \to 0$, where $\beta\hat G({\bf 0})\equiv \partial m
/ \partial h= \beta/[c_0(1-z)]$. Specifically, one has $2d\xi^2=z/(1-z)$ for the
simple  hypercubic  lattice.  Therefore, for   a  given value  of  the
crystal  field $\tilde  \Delta$, the  condition $z=1$  gives  the locus of
diverging   correlation length   and  diverging susceptibility  in the
$(\tilde J,m)$ plane.  This defines a spinodal surface in the $(\tilde
J,\tilde  \Delta,m)$    space; in   particular,   $z(\tilde J,\tilde
\Delta,m=0)=1$  corresponds to the  $\lambda$-line in  the  region of the $h=0$
phase diagram where the transition is continuous.

In  order to  determine the two  unknown  functions $c_0$  and $z$, we
impose thermodynamic self-consistency. To this end, we consider
the change in the  free energy associated  to infinitesimal changes in
$\tilde J,\tilde \Delta$ and $\tilde h$:

\be
\delta \tilde{\cal F}=-\delta \tilde J \sum_{<ij>} <S_{i} S_{j}> +\delta \tilde \Delta
\sum_i <S_i^2>-\delta \tilde h\sum_i<S_i>  \ .
\ee
In terms of the pair correlation function, this gives

\be
\delta \tilde{\cal F}/N=-\frac {1}{2} [G({\bf r}={\bf
e})+m^2] \delta \lambda+ [G({\bf r}={\bf 0})+m^2]\delta \tilde \Delta- m \delta \tilde h 
\ee
where $N$ is the  number of lattice sites  and the coordination number
$c$  has  been  adsorbed in   the  new inverse temperature   variable
$\lambda=c\tilde J$.  The corresponding change in the Gibbs free energy is
 
\be
\delta \tilde{\cal G}/N=-\frac{1}{2}  [G({\bf r=e})+m^2]\delta \lambda+  [G({\bf
r}={\bf 0})+m^2]\delta \tilde \Delta+\tilde h \delta m  \ .
\ee
On the other hand, from Eq. (5), we have
\be
\frac {\partial ^2 \tilde {\cal G}/N} {\partial m^2}=\hat C({\bf k}={\bf 0})  \ .
\ee
Therefore, in  order to get  the same Gibbs free energy when integrating
with respect to $\lambda,\tilde \Delta$ or $m$, the following Maxwell relations
must be satisfied

\begin{mathletters}
\bea
\frac {\partial  \hat C({\bf k}={\bf 0})}{\partial \lambda}&=&-\frac {1}{2}\frac {\partial ^2 }{\partial m^2}[G({\bf r=e})+m^2] \\
\frac {\partial G({\bf r}={\bf 0})} {\partial \lambda}&=&-\frac {1}{2}\frac {\partial G({\bf
r}={\bf e})} {\partial \tilde \Delta}\\
\frac {\partial  \hat C({\bf k}={\bf 0})}{\partial \tilde \Delta}&=&\frac {\partial ^2 }{\partial m^2}[G({\bf r}={\bf 0})+m^2] \  .
\eea 
\end{mathletters}
Clearly,  only two of  these equations  are  independent and in the
following we shall use Eqs. (15a) and (15b).

Replacing $\tilde \Delta$ by the new variable $\tau= (1+\frac{1}{2}e^{\tilde
\Delta})^{-1}$ which varies from $0$  to $1$,  and inserting the  explicit
expressions of $\hat C({\bf  k})$  at ${\bf  k}={\bf 0}$, and  $G({\bf
r})$ at ${\bf r}={\bf 0}$ and ${\bf r}={\bf e}$ that are obtained from
Eqs.  (8-10), we finally get the two SCOZA  equations

\begin{mathletters}
\bea
\frac{\partial}{\partial \lambda}c_0(1-z)&=&-1-\frac{1}{2}\frac {\partial^2 }{\partial m^2}\frac{P(z)-1}{z c_0}\\
\frac{\partial}{\partial \lambda}\frac{P(z)}{c_0}&=&\frac{1}{2}\tau(1-\tau)\frac {\partial
}{\partial \tau}\frac{P(z)-1}{z c_0}
\eea 
\end{mathletters}
where $P(z) \equiv P({\bf r}={\bf 0},z)$  and the relation $P({\bf r}={\bf
e},z)=(P(z)-1)/z$  has  been  used.    Given the appropriate  boundary
conditions,  integration of    these   coupled  partial   differential
equations (PDE)   in   the two unknown   functions  $c_0(\lambda,\tau,m)$ and
$z(\lambda,\tau,m)$ gives at once the thermodynamics of the Blume-Capel model
in the  whole parameter space.    Indeed, according to Eq.  (13),  the
Gibbs free  energy $\tilde {\cal  G}(\lambda,\tau,m)$ can be  obtained in the
same  run  of integration from  the integral  of $-\frac{1}{2} [G({\bf
r=e})+m^2]$  with   respect  to   $\lambda$  (thanks to   the thermodynamic
consistency, this is equivalent to integration with respect to $\tilde
\Delta$   or to $m$).   At  fixed $\lambda$  and $\tau$, $\tilde  {\cal  G}$ as a
function    of  $m$   has     the same      form  as  in    mean-field
theory\cite{BEG1971}, except  that  the   unstable region  inside  the
spinodal is  excluded. At the critical  temperature, there is a single
minimum at $m=0$ for  $\tau>\tau_t$ or three  minima at  $0$ and $\pm  \Delta m$
(with $\tilde  {\cal G}(\pm  \Delta  m)=\tilde {\cal G}(0)$)  for $\tau<\tau_t$.
This defines the  second-  and first-order  parts of  the $h=0$  phase
diagram, respectively.  The maxima of the spinodal curves in the $T-m$
plane (corresponding to $ \partial^2 {\cal G}/\partial m^2=0$) define the lines of
second-order critical  points,   i.e.,   the $\lambda$-line  for   $\tau>\tau_t$
($m=0$)  and the wing critical  lines for $\tau<\tau_t$ ($m=\pm m_c(\tau)$).  In
the   latter case, the    corrresponding  critical field  is given  by
$\pm h_c=\left.\partial    ({\cal  G}/N)/\partial  m   \right|_{m=\pm m_c}$.   The coordinates
($T_t,\tau_t$) of the TCP can be  determined accurately by observing the
change in the convexity of the spinodal at $m=0$.

From a computational  point of view, the   coupled PDE's, Eqs.   (16),
define an initial value problem. The inverse temperature variable $\lambda$
plays the role of time and the equations describe how $z(\lambda,\tau,m)$ and
$c_0(\lambda,\tau,m)$ propagate forward in time.  We thus need to specify the
initial  condition at $\lambda=0$  and the  boundary  conditions at $\tau=1$,
$\tau=0$, and  $m=\pm 1$ (actually, because of  symmetry, one can restrict
the domain of integration to $m\geq 0$).

The  initial  condition  $J=\lambda=0$ corresponds  to the high-temperature
limit where the  spins are independent.  The  properties of the system
can be calculated exactly and  the correlation functions are  non-zero
only at  ${\bf r}={\bf 0}$.  This implies  that $z=0$ and $\hat C({\bf
0})\equiv \partial {\tilde h}/ \partial m=c_0$.  The inverse susceptibility is readily
obtained from the expression of the magnetization

\be
m=\frac {e^{\tilde h}-e^{-\tilde h}} {e^{ \tilde \Delta}+ e^{\tilde h}+e^{-\tilde
h}} \ ,
\ee
which yields

\be
c_0= \frac {[1-\tau+(m^2-2m^2\tau +\tau^2)^{1/2}]^2}{(1-m^2)[(1-\tau)(m^2-2m^2 \tau
+\tau^2)^{1/2}+(m^2-2m^2 \tau +\tau^2)]} \ .
\ee

The boundary condition at $\tau=1$ is given by the solution of the SCOZA
equation for the  spin-$\frac{1}{2}$ Ising model.  Indeed,  this limit
is  reached when  $\tilde  \Delta \to -\infty$,  and  the $S=0$  state  is thus
completely suppressed.  Eq.  (16b) then shows that $P(z)/c_0=f(m)$ and
setting $\tau=1$ in Eq.  (18) gives $f(m)=1-m^2$.   The equation for the
remaining variable $z(\lambda,m)$ , Eq. (16a), becomes
\be
\frac{1}{1-m^2}\frac{\partial}{\partial \lambda}(1-z)P(z)=-1-\frac{1}{2}\frac {\partial^2 }{\partial m^2}[(1-m^2)\frac{P(z)-1}{z P(z)}]
\ee 
which    is the  SCOZA   equation  for   the  Ising  model studied  in
ref.\cite{DS1996}    (with the  standard  replacement   of lattice-gas
variables  by spin  variables).   In  ref.\cite{DS1996}, the   unknown
function  $c_0$  was determined    by  using the   hard-spin condition
$S_i^2=1$  which   implies  that   $G({\bf  r=0})=P(z)/c_0=1-m^2$  (in
lattice-gas language, this is the so-called {\it core} condition).  Since
the  self-consistency  conditions, Eqs.   (15),  are exact,  it it not
surprising that the same result comes out from our equations when the
state $S=0$ is suppressed.

The second boundary at $\tau=0$ is reached  when $\tilde \Delta \to+ \infty$. The
$S=\pm 1$ states are then suppressed.  This only happens, however, if the
magnetic field $h$ is finite.  For $h  \to \pm \infty$,  one can still have a
non-zero magnetization. As a consequence,  $z$ remains a   non-trivial
function  of temperature   and  magnetization.  The solution   of Eqs.
(16b) is again $P(z)/c_0=f(m)$, and setting  $\tau=0$ in Eq.  (18) gives
$f(m)=m(1-m)$ for $m \geq 0$. This leads to the equation
\be
\frac{1}{m(1-m)}\frac{\partial}{\partial \lambda}(1-z)P(z)=-1-\frac{1}{2}\frac {\partial^2 }{\partial m^2}[m(1-m)\frac{P(z)-1}{z P(z)}]
\ee 
which identifies to Eq. (19) by replacing $m$ by $\frac{1-m}{2}$ and $\lambda$ by
$4\lambda$. Therefore, remarquably, the spinodal for $h \to \pm \infty$ can be deduced from
the spinodal of  the spin-$\frac{1}{2}$ Ising model.   The two maxima  at $m_c=\pm
\frac{1}{2}$  and $T_c=\frac{1}{4}T_c^{Ising}$  correspond  to second-order transitions
which mark the end  of the wing   critical lines for $h_c  \to \pm
\infty$, as illustrated  below.

Finally, the boundary $m=1$ is reached when all spins are in the $S=1$
state.  Since there are no more fluctuations,  one has $\xi=0$ and thus
$z=0$, whereas $G({\bf r}={\bf 0})=<S_i^2>-<S_i>^2=P(z)/c_0=0$  implies
that $c_0 \to \infty$.

The numerical  integration of the PDE's was  performed by using a
finite-difference scheme in which the three variables $\lambda, \tau$ and $m$
are discretized   and the   partial derivatives are    approximated by
finite-difference    representations\cite{PFT1992}.      The     first
derivatives with respect to  $\lambda$ are used to  update $z$ and $c_0$ at
the temperature    step  $n+1$ by  evaluating   the   first and second
derivatives with  respect  to $\tau$  and   $m$ at  the   step $n$.   In
principle,  this is a  straightforward procedure.  There are, however,
two difficulties.  First, the region of integration  is bounded by the
spinodal surface which is not known  in advance.  Secondly, there is a
singular behavior as one approaches the spinodal. To  see this, let us
rewrite the PDE's using as unknown functions $z$  and the new variable
$v=P(z)/c_0$. We get
\begin{mathletters}
\bea
A(z,v) \frac {\partial v} {\partial \lambda}+B(z,v) \frac {\partial z} {\partial
\lambda}&=&-1-\frac {1} {2} \frac {\partial^2} {\partial m^2} (\psi(z)v)\\ 
\frac {\partial v} {\partial \lambda} &=& \frac {1}{2}\tau (1-\tau) \frac {\partial} {\partial\tau}
(\psi (z)v)
\eea
\end{mathletters}
where $A(z,v)=-P(z)(1-z)/v^2$, $B(z,v)=(1/v) \partial [(1-z)P(z)]/ \partial z$ and
$\psi(z)=(P(z)-1)/(zP(z))$.  Eq. (21a)   can  be viewed as  a  nonlinear
diffusion equation   and  $A^{-1}$  plays  the   role of a   diffusion
coefficient that diverges like $(1-z)^{-1}$  when $z \to 1$, namely  on
the spinodal surface. These two   difficulties are already present  in
the equation  for the Ising model,  Eq.(19), but then  the spinodal is
just a line in the $(\lambda,m)$ plane\cite{DS1996}.

We   solved the  first problem   as   follows.  Whenever the  variable
$z_n(\tau,m)$  at  the   temperature   step $n$  enters   the   interval
$(1-\epsilon,1)$, where $\epsilon$  is small  number  (typically, $\epsilon  \leq 10^{-5}-
10^{-6}$), we consider  that the spinodal is  reached and we  stop the
calculation.  The spinodal is then defined by the corresponding values
of $\tau$ and $m$. At the next temperature step,  we use the same values
$z_n$ and $v_n$  to compute the   partial derivatives with respect  to
$\tau$ and $m$ on the  spinodal (in other  words, we locally  ``freeze''
the  values of $z$ and  $v$ inside the  spinodal).  On the other hand,
the  vanishingly  small  values of $A(z,v$)  for  $z   \to 1$ impose  a
dramatic  decrease of  the inverse temperature  spacing  $\Delta\lambda$ as the
spinodal  is  approached.    Indeed, as is   well known\cite{PFT1992},
solving a diffusion  equation by an  {\it explicit} method requires to
keep the ``time"  step below a certain value  which is proportional to
$D^{-1}$,  where   $D$     is   the  diffusion      coefficient.    In
ref.\cite{DS1996}, this problem was avoided  by using an {\it implicit
} method which is  inconditionally stable.   Unfortunately, it is  not
easy to generalize this procedure to a system  of nonlinear PDE's like
Eqs.   (21) and we  had to keep  a simple explicit  algorithm.  In the
high-temperature region, we typically   adopted the spacings $\Delta  m=\Delta
\tau=10^{-2}$ and $\Delta \lambda=10^{-4}$.  In the vicinity of the $\lambda$-line and
in the tricritical region, $\Delta \tau$ was set  at $2.10^{-3}$ and $\Delta \lambda$
was gradually decreased down to  $10^{-7}$.  This allowed to determine
the critical parameters with excellent accuracy.  For instance, in the
limit $\tau\to 1$, we found $\tilde J_c=0.22125$ for the inverse critical
temperature  of the  Ising model  on  the simple cubic  lattice.  This
corresponds to   $\tilde J_c^{LG}=4\tilde  J_c=0.88500$  for  the n.n.
lattice-gas,    in perfect  agreement  with   the   value obtained  in
ref.\cite{DS1996} (this is also within  $0.2 \%$ of the  best-estimate
result\cite{BLH1995}).    When  higher   accuracy was  required,   for
instance to determine the  asymptotic critical exponents, $\Delta  \lambda$ was
further decreased to  $10^{-9}$.  The integration was  usually carried
down to $k_BT/Jc  \approx 0.18$ (below this  value, the spinodal  lines in
the  vicinity of $\tau=1$  are so  close to  the  boundary $m=1$ that we
could not integrate  the  equations  with sufficient accuracy    while
keeping a reasonable spacing $\Delta m$;  indeed, decreasing $\Delta m$ forces
to  decrease  also $\Delta      \lambda$    in order   to    avoid    numerical
instabilities\cite{PFT1992}).

Before    presenting  the  numerical  results, it    is instructive to
consider the high-temperature series expansion   of the solution and
compare it to the exact results.   Since $z \to 0$ for  $\lambda \to 0$, one can
replace  the  Green's function $P(z)$ by  its   expansion in powers of
$z$. We  then expand $z$  and $c_0$ as double  series in $\lambda$ and $m$,
$z(\lambda,\tau,m)=\sum_{pq}     z_{pq}(\tau)     \lambda   ^{p}      m^{2q}$      and
$c_0(\lambda,\tau,m)=\sum_{pq} c^0_{pq}(\tau)\lambda  ^{p} m^{2q}$.  The  coefficients
$z_{pq}(\tau)$ and $c^0_{pq}(\tau)$ are polynomials of $\tau$ that satisfy a
system of linear algebraic equations at each order in $\lambda$ and $m$. In
the       case     of      the         fcc        lattice     ($c=12$,
$P(z)=1+z^2/12+z^3/36+5z^4/192+5z^5/288+...$)  for    which  extensive
series expansions have been derived by Saul {\it et al.}\cite{FG1972},
the  results for the  zero-field ordering susceptibility $\chi_0=\sum_{\bf r}G({\bf
r};m=0)=
\left.  \partial m/  \partial h\right|_{h=0}$    and   the second moment of    the
correlation function $\mu_2=\sum_{\bf r}r^2G({\bf r};m=0)=c\xi^2\chi_0$ are
\bea
k_BT \chi_0 = &\tau& +12\tau^2\tilde J
+6(\tau^2+21\tau^3)\tilde J^2+2(\tau^2+78\tau^3+621\tau^4)\tilde J^3+\frac{1}{2}(\tau^2+234\tau^3
+5115\tau^4\nonumber\\
&+&23778\tau^5)\tilde J^4+
\frac{1}{10}(\tau^2+612\tau^3+31851\tau^4+342690\tau^5+1122462\tau^6)\tilde J^5\nonumber\\
&+&\ldots
\eea
and
\bea
\mu_2=&12&\tau^2\tilde J+288\tau^3\tilde J^2+2(\tau^2+66\tau^3+2385\tau^4)\tilde
J^3+\frac{1}{2}(240\tau^3+10080\tau^4+133488\tau^5)\tilde J^4\nonumber\\
&+&\frac{1}{10}(\tau^2+492\tau^3+50931\tau^4+1156410 \tau^5+8474742\tau^6)\tilde
J^5+ \ldots
\eea
Comparison   with  the   exact   series expansions   shows   that  the
coefficients  in both expressions are  exact  through order $(\beta J)^4$
while  higher-order terms do not  deviate significantly from the exact
ones. This  is similar  to the case   of the  spin-$\frac{1}{2}$ Ising
model  and we  take   this result as  a  strong  indication   that the
numerical  predictions of the SCOZA should  be very close to the exact
solution of the model.

To close this section,  let us note that the  present theory does  not
provide a {\it complete} description of the  system. In particular, it
does not give any information concerning the two correlation functions
$G_{ij}^{S^2S^2}=<S^2_iS^2_j>-<S^2_i><S^2_j>$                      and
$G_{ij}^{SS^2}=<S_iS^2_j>-<S_i><S^2_j>$.  In  order to determine these
functions,  one  needs to  introduce   a set  of three  different direct
correlation   functions.  This implies to    perform a double Legendre
transform that  defines a new Gibbs  free energy which is a  function of
$m$ and $x$,  the concentration order   parameter, instead of $m$  and
$\Delta$. The  main interest of this  alternative theory is that it allows
to study the general  spin-1 Hamiltonian with   $K\neq 0$ and  $L\neq
0$. On the other hand,  the numerical solution  is more difficult as
one has to solve  three coupled PDE's instead  of two. Details on the
derivation of these equations are given in  Appendix A.

\section{Results}

In this section we concentrate  on the SCOZA numerical predictions for
the  phase  boundaries.  These are  nonuniversal  properties which are
lattice-dependent. If not stated otherwise,  the results presented here
correspond to the  simple cubic  lattice  for which no systematic
study has been performed in the literature.

The overall shape of the spinodal surface  in the $(T,\tau,m)$ space and
the vicinity of  the TCP are depicted  in Figs. 1 and 2, respectively.
We clearly  see  the  evolution from the single curve  at $\tau=1$  (the
spinodal of the spin-$\frac{1}{2}$ model) which has a maximum at $m=0$
to the  two symmetrical  curves at $\tau=0$   with maxima located  at $m_c=\pm
\frac{1}{2}$, marking the end of the wing critical lines.

The $T_c(\tau)$,  $T_c(\Delta)$, and $T_c(x)$  phase diagrams  in zero field
are shown in Figs.  3-5.  Second- and first-order phase boundaries are
shown  as full and dashed lines,  respectively.   The curves are quite
similar  to   those obtained   by  series  expansions\cite{FG1972} and
Monte-Carlo  simulations\cite{JL1980}   for  the   fcc   lattice.   In
particular, we see that the slope of the phase boundary across the TCP
is  finite   and     continuous  in  both   $T_c(\tau)$    and $T_c(\Delta)$
(specifically, we find $ (T_t/Jc)\left.  \partial\Delta /\partial T
\right|_{T_t}=-0.045$);  the   $T_c(\Delta)$   phase boundary is  slightly
concave upward   just below the   TCP,  and the  $\lambda$-line appears  to
extrapolate into the interior of the two-phase  region in the $T_c(x)$
phase  diagram (a continuous slope,  however, cannot be strictly ruled
out by our calculations). As is well known, the  slope of the $\lambda$-line
and the slope of the coexistence curve on the $^3He$-rich side are not
the   same    experimentally.   This        is also   predicted     by
renormalization-group   analysis\cite{RW1972},     in   contrast  with
mean-field theory\cite{BEG1971}.

The accuracy of our calculation for  the $\lambda$-line in the $\Delta-T$ plane
can be checked for  the special value   $\tilde \Delta=\ln 2$ for  which a
careful  Monte-Carlo calculation   and   finite-size study   has  been
performed by Bl\"{o}te {\it et al.}\cite{BLH1995}.  Our prediction for
the inverse critical temperature  $\tilde J_c=J/(k_BT_c)=0.3924$ is in
excellent  agreement their  estimate $\tilde  J_c=0.3934224(10)$.  The
accuracy of the theory is thus the  same as for the spin-$\frac{1}{2}$
Ising model\cite{DS1996}.

As noted earlier, the TCP corresponds to the point where the convexity
of  the spinodal  in   the $T-\tau$ plane    changes at $m=0$.   This is
illustrated in Fig.   6 which shows  the temperature dependence of the
order parameter and the  spinodal lines in  the tricritical region  on
the first-order side  of the phase  boundary (observe that $\Delta m(\tau)$,
the discontinuity in the order parameter  across the first-order phase
boundary, moves  away  from  the  spinodal as   $\tau$ decreases).   The
coordinates of the  tricritical point are  $k_BT_t/J=1.4160 \pm0.0040$,
$\tau_t=0.2114\pm 0.0010$ ($\Delta_t/J=2.8457$), $x_t=0.655\pm 0.006$, where the
uncertainties reflect   the finite size  of  the  grid  spacings.  The
predictions  for $T_t$ and $\Delta_t$  are in excellent agreement with the
recent       Monte-Carlo     estimates   of      Deserno\cite{JL1980}:
$k_BT_t/J=1.4182\pm  0.0055$,  $ \Delta_t/J=2.8448\pm 0.0003$  (these numbers,
however, are different  from  those quoted in Ref.\cite{HB1998}  which
locate   the    TCP   near $k_BT_t/J=1.3900$,  $    \Delta_t/J=2.849$, and
$x_t=0.61$; if  these (unpublished) results are  correct, our value of
$x_t$   is overestimated and  too close  to the mean-field prediction,
$x_t^{MF}=2/3$).    Similarly,  for    the   fcc   lattice,    we find
$k_BT_t/J=3.1116  \pm0.0090$,  $\tau_t=0.2454\pm  0.0010$ ($\Delta_t/J=5.6520$),
$x_t=0.658\pm   0.006$,  which  we   may compare  with  the  Monte-Carlo
estimates  of Jain and  Landau\cite{JL1980}  $k_BT_t/J=3.072\pm  0.024$,
$\Delta_t/J=5.652\pm 0.048$, $ x_t=0.56\pm 0.02$ (note that our value of $x_t$
is in much better agreement with the series expansion estimate of Saul
{\it et  al.}\cite{FG1972},  $x_t=0.665^{+0.005}_{-0.015}$; obviously,
further  work is  needed to  locate  precisely the  TCP  in the  $x-T$
plane).  Finally,   our   predictions   for  the   bcc    lattice  are
$k_BT_t/J=2.0264 \pm0.0060$, $\tau_t=0.2354\pm  0.0010$   ($\Delta_t/J=3.7918$),
$x_t=0.656\pm  0.006$  (to our knowledge,   this  lattice has  only been
studied by real-space renormalization-group methods\cite{BW1976} which
do not predict accurately the location of the tricritical point).

At the TCP, the $\lambda$-line bifurcates into two symetrical wing critical
lines. The projections of the wing boundaries onto the $\Delta-T$, $\Delta-h$,
and     $T-h$    planes   are   shown  in       Fig.    7.  Mean-field
theory\cite{BEG1971} predicts that the critical  field $h_c$ should go
to infinity at $k_BT_c/Jc=\frac{1}{4}$. We  clearly see in Fig. 7 that
this value is overestimated. In  fact, as  noted earlier, the  present
theory predicts that $h_c \to\pm \infty$ at $k_BT_c/Jc=
\frac{1}{4}k_BT_c^{Ising}/Jc=0.188$. For the fcc lattice, this yields
$k_BT_c/Jc=0.204$,  which  is consistent  with the  value  that can be
extracted  from     the   Monte-Carlo   simulations   of     Jain  and
Landau\cite{JL1980}.

\section{Asymptotic behavior in the critical  and tricritical regions}

As  mentioned  in  the     Introduction,  the  SCOZA  for   the    3-d
spin-$\frac{1}{2}$  Ising model has  a nontrivial  scaling behavior in
the critical region\cite{DS1996,HPS2000}.  Above $T_c$, the asymptotic
behavior is the same  as in the  mean-spherical approximation and  the
exponents are those of  the spherical model.  This  spherical scaling,
however, is  detectable only   in a  very narrow neighborhood   of the
critical  point, and the  effective SCOZA exponents   are close to the
true  Ising ones down  to   reduced temperatures of around  $10^{-2}$.
Below $T_c$,  the scaling is neither  spherical nor classical with two
scaling   functions   instead  of  one\cite{HPS2000}.     Despite this
shortcoming, the zero-field magnetization is very well described, with
an  asymptotic  exponent $\beta=7/20=0.35$ which  is   close to the exact
value $\beta  \approx 0.33$.  It is thus  interesting to also  investigate the
critical behavior   of  our  SCOZA  equations  in  the  critical   and
tricritical regions of the 3-d BC model.

We   first     consider the  behavior   of   the   zero-field ordering
susceptibility  $\chi_0$ as $T \to T_c(\tau)$  along paths of constant $\tau$
for $\tau\geq  \tau_t$. Accurate evaluations were relatively straightforward
to perform in  the disordered phase:  we  only had to  gradually
decrease the spacing $\Delta  \lambda$ as discussed earlier.   Fig.  8 shows  a
log-log plot of $k_BT\chi_0$  as a function  of the  reduced temperature
$t=1-\frac{T_c}{T}$ together with the corresponding effective exponent
$\gamma_{eff}$ defined as the local  slope $\partial\log(k_BT\chi_0)/ \partial \log  t$.
In the region $1.0\geq \tau>0.25$,  it can be seen that
each $\gamma_{eff}(\tau)$ reaches the value $2$ for $t \sim 10^{-5}$ as in the
case  of the  spin-$\frac{1}{2}$  model, showing  that the  asymptotic
spherical behavior is universal.  This is no  more true when one moves
further away from  $T_c$.   However, in  the  range  $t\geq 10^{-2}$,  a
quasi-universal behavior is still  observed for $1.0\geq \tau>0.6$, and  the
critical behavior is governed by an effective  exponent which is close
to  the exact Ising value   $\gamma\approx 1.24$. On the   other hand, as  $\tau$
approaches its tricritical   value  $\tau_t=0.211$, there  is  an abrupt
crossover to another  behavior   which is  governed by  the   exponent
$\gamma_{eff}  \approx 1$  over  a wide range   of temperatures.  There is good
numerical evidence  that   $\gamma_{eff}$ reaches $1$   asymptotically  at
$\tau=\tau_t$.

For subcritical temperatures, it was more difficult to obtain accurate
results in  the vicinity  of $T_c$ because  of our  use of an explicit
method to  integrate  the PDE's.  Accordingly, we   were  only able to
explore  the critical behavior  in a  restricted range of temperatures
$t=1-\frac{T}{T_c}$.  Log-log  plots of the temperature  dependence of
the  order parameter $m$ are  shown  in Figs. 9  and  10.  Despite the
limited range, it appears from Fig.  9 that in the second-order region
well above the  crossover to tricritical  behavior, the slope  of each
curve has a  common asymptotic  limit  which corresponds to the  SCOZA
prediction  for the Ising model,  $\beta^{Ising}_{scoza}=7/20$.   Fig. 10  shows that for
smaller values of $\tau$, one needs to go  closer to $T_c$ to reach this
asymptotic universal regime.  Again, we observe an abrupt crossover to
another  behavior as one enters  the  tricritical region.  Our results
are consistent with the asymptotic exponent $\beta_t=1/4$ for $\tau =\tau_t$.

Finally, we analyze   the shape of  the  wing critical boundary  as it
approaches the TCP.  Figs. 11 shows the log-log  plots of the critical
field $h_c$ and the  magnetisation $m_c$ as a  function of the reduced
temperature $1-\frac{T_c}{T_t}$.  Our numerical  data are
consistent with  asymptotic power-law  behaviors governed  by the  exponents
$5/2$ for $h_c$ and $1/2$ for $m_c$.

All   the  above numerical results   strongly  suggest that our theory
describes the  whole  critical  portion of the   phase
boundary  by  the   same exponents  as    the SCOZA equation   for the
spin-$\frac{1}{2}$ Ising  model\cite{DS1996,HPS2000} and that near the
TCP there is   a  crossover to  a tricritical  behavior  described  by
mean-field  exponents.   This   is  supported, and    can be   further
rationalized, by  considering a heuristic     scaling analysis of   the
coupled SCOZA  PDE's, Eqs.  (16).  The  argument  is summarized below  and
detailed in Appendix B.

Let us assume that  close to the TCP, the singular part
of the Gibbs free energy can be written as

\begin{mathletters}
\bea
{\cal G}_{sing} &\approx& |t|^{2-\alpha}{\cal G}_{\pm }(\frac {g}
{|t|^{\phi}},\frac{m} {|t|^\beta}) \\
&\approx& |g|^{2-\alpha_t}{\cal G}^t_\pm (\frac {t}
{|g|^{\phi_t}},\frac{m} {|g|^{\beta_t}})
\eea
\end{mathletters} 
where we have introduced   the two scaling fields  $t=(T-T_t)/T_t$ and
$g=(\tau-\tau_t)/   \tau_t-at$   where  $(\tau_t/T_t)a=\left.   \partial\tau_{\lambda} /\partial
T\right|_{T_t}>0$ is the slope of the $\lambda$-line at the TCP (this is also
the slope  of the triple  line below $T_t$, as we   have seen that the
slope of  the  phase boundary is  finite and  continuous  at the TCP).
Eqs. (24)   have  the  form   of   the  standard  tricritical  scaling
hypothesis\cite{RW1972,LS1984}, except  that we use  the magnetization
$m$ instead of the magnetic field  $h$ as variable.  ${\cal G}_{(\pm) }$
and ${\cal G}^t_{(\pm) }$ are the scaling functions, where the subscript
$(\pm)$ represents the  sign of $t$, i.e.,  denotes when the temperature
is above or below the  tricritical temperature $T_t$.  When $|t| \to 0$
with $g= 0$, the TCP is  approached tangentially to the phase boundary
in the symmetry plane $h=0$ (and Eq. (24a) is then the convenient form
of  the scaling hypothesis),  whereas   the TCP  is approached with  a
finite angle with  the phase boundary when  $g\neq 0$ (and  Eq. (24b) is
the  convenient scaling form).  This  defines  two  sets of  exponents
$(\alpha,\phi,\beta)$  and   $(\alpha_t,\phi_t,\beta_t$)   that    are   related through
$2-\alpha_t=(2-\alpha)/ \phi $, $\phi_t=1/ \phi $ and $\beta_t=\beta/ \phi$.

Since SCOZA is  thermodynamically self-consistent, the scaling
behavior  of the Gibbs  free energy near the  TCP  is inherited by its
various derivatives.  In  particular, the asymptotic behavior   of the
magnetic field $h=\partial({\cal G}/N)/ \partial m$ is

\be
h \approx   |t|^{2-\alpha -\beta}\frac{\partial}{\partial   v}{\cal G }_{\pm  }(u,v)\approx
|g|^{2-\alpha_t-\beta_t}\frac{\partial}{\partial v_t}{\cal G }^t_{\pm }(u_t,v_t)
\ee
where      $u=g/|t|^{\phi}$,     $v=m/|t|^{\beta}$     ($u_t=t/|g|^{\phi_t}$,
$v_t=m/|g|^{\beta_t}$),   and   the   inverse   ordering   susceptibility
$\chi_0^{-1}=\partial^2( {\cal  G}/N)/ \partial m^2$ and  the  singular part  of the
$^4He$ concentration  order  parameter $\rho=1-x=\partial( {\cal  G}/N)/ \partial\Delta$
obey (up to irrelevant multiplying factors)

\be
\chi_0^{-1} \approx   |t|^{\gamma}\frac{\partial^2}{\partial   v^2}{\cal G }_{\pm  }(u,v)\approx
|g|^{\gamma_t}\frac{\partial^2}{\partial v_t^2}{\cal G }^t_{\pm }(u_t,v_t)
\ee
and 
\be
\rho_{sing}\approx - |t|^{2-\alpha-\phi}\frac{\partial}{\partial u}{\cal G }_{\pm }(u,v)\approx - |g|^{2-\alpha_t-\phi_t}\frac{\partial}{\partial u_t}{\cal G }^t_{\pm }(u_t,v_t)
\ee
where $\gamma=2-\alpha-2\beta$  (resp.
$\gamma_t=2-\alpha_t-2\beta_t$). Note also that because $m$ is used in Eqs. (24)
instead of $h$, the zero-field magnetization  is solution of
the implicit equation $h=\partial ({\cal G}/N)/  \partial m=0$. This implies that
the singular part of this quantity near the TCP obeys
$m_{sing} \approx |t|^{\beta}{\cal M}_{\pm}(g/|t|^{\phi}) \approx |g|^{\beta_t}{\cal
M}^t_{\pm}(g/|t|^{\phi_t})$.

From the numerical  results shown in  Figs.   (8-11), it appears  that
$\gamma_t\approx 1$ (see   the curve $\tau=0.22$  in  the lower part  of  Fig.8),
$\beta_t\approx 1/4$ (see the curve $\tau=0.21$ in Fig.   10), $2-\alpha -\beta \approx 5/2$
and $\beta \approx 1/2$ (upper and lower parts of Fig.11).  (We have also good
numerical evidence that    $\beta \approx 1/2$  from  a  log-log plot  of  the
discontinuity  of the  zero-field    magnetization as  a   function of
$(T_t-T)/T_t$ across   the  first-order  phase boundary.)    All these
exponents  have their classical values,  and from the scaling relation
$\alpha+2\beta +\gamma=2$ (resp. $\alpha_t+2\beta_t +\gamma_t=2$), we deduce that $\gamma=2$ and
$\alpha_t=\frac{1}{2}$.  These values are all  consistent with a crossover
exponent  $\phi=2$. 

We  show in  Appendix B   that  the  scaling  ansatz,  Eqs.  (24)   or
Eqs. (26-27), is compatible with the asymptotic behavior of the PDE's,
Eqs (16), in  the tricritical region.  This  analysis indicates that a
non-trivial scaling is found when the exponents obey the two relations
$\gamma=\phi$  and  $\gamma=4\beta$, which  are  satisfied by the classical values.
Morever, one  finds   that  the scaling  function of    the zero-field
susceptibility above $T_t$ obeys an equation which is quite similar to
the  asymptotic SCOZA equation   studied in Ref.\cite{HPS2000} for the
spin-$\frac{1}{2}$ model.  It     can be inferred   that  the critical
behavior of the present theory along the $\lambda$-line  is the same as the
SCOZA prediction  for the Ising  model.  This  is  consistent with the
exponent $\dot{\beta}=7/20$ which is observed  numerically in Fig.  9 along the
high-temperature part of  the $\lambda-$line.  This universality appears to
hold along the wing critical lines too since the boundary condition to
Eqs.  (16) at the end of these lines (for $h_c \to  \pm \infty$) is again the
SCOZA PDE for the Ising model, as explained in section 2.

\section{Conclusion}

The  present study shows  that a  thermodynamically self-consistent OZ
approximation  provides a very good  description of  the properties of
the $3$-d Blume-Capel model in all parts of the phase diagram. Like in
the  case of  the Ising  model, non-universal  properties such as  the
shape  of the  phase boundaries  and the  location of  the tricritical
point are predicted with remarkable accuracy.  Moreover, there is good
numerical and analytical evidence  that the SCOZA correctly predicts a
universal critical behavior along the  $\lambda$-line and the wing critical
lines (with a zero-field   magnetization exponent $0.35$ that is  very
close to the true Ising value), as well as  a crossover to tricritical
behavior governed by classical exponents.  Therefore, the SCOZA proves
to   be a  powerful  tool for   studying   spin systems  which exhibit
first-order   and/or  continuous transitions.   This  is  confirmed by
further   work  on  the    ferromagnetic  spin-$3/2$\cite{G2000}   and
Potts\cite{GRT2000} models.
\newpage

\appendix

\section{}

In this Appendix, we derive the SCOZA equations for the
the most general spin-1 Hamiltonian with n.n. couplings,
\be
{\cal H} = {\cal H}_{BC} -K \sum_{<ij>}S_i^2S_j^2-L
\sum_{<ij>}S_iS_j(S_i+S_j) \ ,
\ee
which is a model   for ternary mixtures\cite{MB1974}. The  solution of
these  equations also provides  a  complete description of the pair
correlation functions of  the Blume-Capel model.  These  functions can
be generated by introducing  site-dependent fields $h_i$ and $\Delta_i$ in
the Hamiltonian (A1), which yields

\begin{mathletters}  
\bea
G_{ij}^{SS}&=&<S_iS_j>-<S_i><S_j>= -\frac{\partial ^2 \tilde {\cal F}}{\partial\tilde h_i \partial \tilde h_j}\\ 
G_{ij}^{SS^2}&=&<S_iS_j^2>-<S_i><S_j^2>=\frac{\partial ^2 \tilde {\cal F}}{\partial\tilde h_i \partial \tilde \Delta_j}\\
G_{ij}^{S^2S^2}&=&<S_i^2S_j^2>-<S_i^2><S_j^2>= -\frac{\partial ^2 \tilde
{\cal F}}{\partial\tilde \Delta_i \partial \tilde \Delta_j} \ .
\eea 
\end{mathletters}
We then perform a double Legendre transform that takes the fields $h_i$
and     $\Delta_i$  into    $m_i$    and    $x_i$,  respectively,    where
$x_i=1-<S_i^2>=1-\partial  {\cal  F}/ \partial  \Delta_i$.   This  defines  a  Gibbs
free energy  ${\cal G  }(T,\{m_i\},\{x_i\})={\cal  F } +  \sum_i h_i m_i
-\sum_i \Delta_i (1-x_i)$ which satisfies $h_i= \partial {\cal G }/\partial m_i$ and
$\Delta_i=\partial {\cal G } /\partial x_i$. ${\cal G  }$ is the generating functional
of the direct correlation functions,
\begin{mathletters}
\bea
C_{ij}^{SS}=\frac{\partial ^2 \tilde {\cal G}}{\partial m_i \partial m_j}\\ 
C_{ij}^{SS^2}=-\frac{\partial ^2 \tilde {\cal G}}{\partial m_i \partial x_j}\\ 
C_{ij}^{S^2S^2}= \frac{\partial ^2 \tilde {\cal G}}{\partial x_i \partial x_j} \ ,
\eea  
\end{mathletters}
that  are related  to the   $G$'s via a set of Ornstein-Zernike
equations. In the limit of uniform fields,  these equations in Fourier
space  take   the form   ${\underline{\underline {\hat {\bf C}}}}({\bf
k}){\underline{\underline  {\hat    {\bf  G}}}}({\bf     k})=1$,   where
${\underline{\underline     {\hat    {\bf    G}}}}({\bf  k})$        and
${\underline{\underline {\hat  {\bf C}}}}({\bf k})$  are symmetrical square
matrices. This readily yields

\begin{mathletters}
\bea
\hat G^{SS}({\bf k})&=&\frac { \hat C^{S^2S^2}({\bf k})} {\hat C^{S^2S^2}({\bf k})
\hat C^{SS}({\bf k}) -  {\hat C^{SS^2}({\bf k})}^2 }\\
\hat G^{S^2S^2}({\bf k})&=&\frac { \hat C^{SS}({\bf k})} {\hat C^{S^2S^2}({\bf k})
\hat C^{SS}({\bf k}) -  {\hat C^{SS^2}({\bf k})}^2 }\\
\hat G^{SS^2}({\bf k})&=&\frac {- \hat C^{SS^2}({\bf k})} {\hat C^{S^2S^2}({\bf k})
\hat C^{SS}({\bf k}) -  {\hat C^{SS^2}({\bf k})}^2 } \ .
\eea
\end{mathletters}
 
We now assume that the range of the direct correlation
functions is limited to n.n. separation, i.e.,

\begin{mathletters}
\bea
{\hat C}^{SS} ({\bf k})&=&c^{SS}_0[1- \zeta_{SS} {\hat \lambda} ({\bf k})]\\
{\hat C}^{SS^2} ({\bf k})&=&c^{SS^2}_0 [1-\zeta_{SS^2}{\hat  \lambda} ({\bf k})]\\
{\hat C}^{S^2S^2} ({\bf k})&=&c^{S^2S^2}_0 [1-\zeta_{S^2S^2} {\hat \lambda} ({\bf k})]
\eea
\end{mathletters}
where the $c_0$'s and the $\zeta$'s are functions of  $T,
m$ and  $x$ to be determined. This  fixes  the  form of the correlation functions in
${\bf r}$-space,  and after some calculations we find

\begin{mathletters}
\bea
G^{SS} ({\bf r})&=&G^{SS} ({\bf r}={\bf 0}) \frac {(z_1-\zeta_{S^2S^2})
P(z_1,{\bf r})-(z_2-\zeta_{S^2S^2}) P(z_2,{\bf r})} {(z_1-\zeta_{S^2S^2})
P(z_1)-(z_2-\zeta_{S^2S^2}) P(z_2)}\\
G^{SS^2} ({\bf r})&=&G^{SS^2}({\bf r}={\bf 0})\frac {(z_1-\zeta_{SS^2})
P(z_1,{\bf r})-(z_2-\zeta_{SS^2}) P(z_2,{\bf r})} {(z_1-\zeta_{SS^2})
P(z_1)-(z_2-\zeta_{SS^2}) P(z_2)} \\
G^{S^2S^2} ({\bf r})&=&G^{S^2S^2}({\bf r}={\bf 0}) \frac {(z_1-\zeta_{SS})
P(z_1,{\bf r})-(z_2-\zeta_{SS}) P(z_2,{\bf r})} {(z_1-\zeta_{SS})
P(z_1)-(z_2-\zeta_{SS}) P(z_2)}
\eea
\end{mathletters}
where $P(z,{\bf r})$ is the lattice Green's function defined by
Eq. (10), and $z_1, z_2$ are related to the  $c_0$'s and the $\zeta$'s
via the relations 

\begin{mathletters}
\bea
\zeta_{SS} + \zeta_{S^2S^2}&=&2 R ^2 \zeta_{SS^2} +(1-R^2) (z_1+z_2)\\
\zeta_{SS} \   \zeta_{S^2S^2}&=& R^2\zeta_{SS^2}^2+(1-R^2) z_1z_2
\eea
\end{mathletters}
and 
\be
\frac{R}{R_0} =\frac {[z_1 P(z_1)-z_2
P(z_2)-\zeta_{SS}(P(z_1)-P(z_2))]^{1/2}[z_1 P(z_1)-z_2
P(z_2)-\zeta_{S^2S^2}(P(z_1)- P(z_2))]^{1/2}} {z_1 P(z_1)-z_2 P(z_2)-\zeta_{SS^2}
(P(z_1)-P(z_2))}
\ee
where  $R=c^{SS^2}_0/(c^{SS}_0 c^{S^2S^2}_0)^{1/2}$  and $R_0$ is  the
high-temperature  limit  of $R$  which  can be   calculated exactly, as
explained below. Note that in Eqs. (A6) we have eliminated the $c_0$'s
to introduce the on-site values of the correlation functions which are
simple functions of  the order parameters $m$  and $x$.  Indeed, since
$S_i$ can only  take the values $0,\pm  1$, one  has $<S_i^3>=<S_i>$ and
$<S_i^4>=<S_i^2>$, so that
\begin{mathletters}
\bea
G^{SS}({\bf r}={\bf 0})& =& 1-x-m^2 \\
G^{SS^2} ({\bf r = 0})&=& mx \\
G^{S^2S^2}({\bf r}={\bf 0})& =& x(1-x) \ . 
\eea
\end{mathletters}
In terms of these  variables, one also has 

\begin{mathletters}
\bea
\hat C^{SS}({\bf k}&=&{\bf 0})=\frac {1-\zeta_{SS}} {(1-R ^2)(1-x-m^2)} \
 \frac{z_1 P(z_1)-z_2 P(z_2)-\zeta_{S^2S^2}[P(z_1)- P(z_2)]} {z_1-z_2}\\
\hat C^{SS^2}({\bf k}&=&{\bf 0})=-\frac {(1-\zeta_{SS^2})R^2} {(1-R ^2)xm} \
 \frac{z_1 P(z_1)-z_2 P(z_2)-\zeta_{SS^2}[P(z_1)- P(z_2)]} {z_1-z_2}\\
\hat C^{S^2S^2}({\bf k}&=&{\bf 0})=\frac {1-\zeta_{S^2S^2}} {(1-R ^2)x(1-x)} \
 \frac{z_1 P(z_1)-z_2 P(z_2)-\zeta_{SS}[P(z_1)- P(z_2)]} {z_1-z_2} \ .
\eea 
\end{mathletters}

It remains three unknown functions to be determined,  and it is convenient
to choose $z_1,   z_2$ and $R$   and to  use   Eqs. (A7) and (A8)   to
calculate the  $\zeta$'s.  The  three additional  equations  that we need   are
obtained by imposing thermodynamic self-consistency.  On the one hand,
the enthalpy is given by

\be
\frac{\partial \tilde {\cal G}/N}{\partial \beta}=-\frac{Jc}{2}[G^{SS}({\bf r}={\bf
e})+m^2]-\frac{Kc}{2}[G^{S^2S^2}({\bf r}={\bf e})+(1-x)^2]-Lc[G^{SS^2}({\bf r}={\bf e})+m(1-x)] \ .
\ee
On the other hand, we have from Eqs. (A3)

\begin{mathletters}
\bea
\hat C^{SS}({\bf k}&=&{\bf 0})=\frac{\partial ^2 \tilde {\cal G}}{\partial m^2}\\ 
\hat C^{SS^2}({\bf k}&=&{\bf 0})=-\frac{\partial ^2 \tilde {\cal G}}{\partial m \partial x}\\ 
\hat C^{S^2S^2}({\bf k}&=&{\bf 0})= \frac{\partial ^2 \tilde {\cal G}}{\partial
x^2} \ .
\eea  
\end{mathletters}
This yields the three Maxwell equations

\begin{mathletters}
\bea
\frac {\partial \hat C^{SS}({\bf k}={\bf 0})} {\partial \lambda }&=&-1-\frac{1}{2}\frac{\partial^2 [G^{SS}({\bf r}={\bf e})+\alpha_1G^{S^2S^2}({\bf r}={\bf e})+2\alpha_2G^{SS^2}({\bf r}={\bf e})]} {\partial m^2}\\
\frac{\partial\hat C^{SS^2}({\bf k}={\bf 0})} {\partial
\lambda}&=&-\alpha_2+\frac{1}{2}\frac{\partial^2[ G^{SS}({\bf r}={\bf e})+\alpha_1G^{S^2S^2}({\bf r}={\bf e})+2\alpha_2G^{SS^2}({\bf r}={\bf e})]}
{\partial m \partial x} \\
\frac {\partial\hat C^{S^2S^2}({\bf k}={\bf 0})} {\partial \lambda}&=&-\alpha_1-\frac{1}{2}\frac {\partial^2 [G^{SS}({\bf r}={\bf e})+\alpha_1G^{S^2S^2}({\bf r}={\bf e})+2\alpha_2G^{SS^2}({\bf r}={\bf e})]} {\partial x^2} 
\eea 
\end{mathletters}
where  $\lambda=\beta J c$, $\alpha_1=K/J$, and $\alpha_2=L/J$.

These equations, together with Eqs. (A7) and (A8), constitute a set of
three PDE's in the unknown functions $z_1(\lambda,m,x)$, $z_2(\lambda,m,x)$, and
$R(\lambda,m,x)$, whose solution encodes the full thermodynamics of the
model Hamiltonian (A1). The initial conditions for $J=K=L=0$ are easily 
obtained since the correlation functions are then non-zero at ${\bf
r}={\bf 0}$ only. One has $z_1=z_2=0$ and from  Eqs. (A4) and (A9)

\be
R_0=-\frac{mx}{[x(1-x)(1-x-m^2)]^{1/2}} \ .
\ee

It sould be noticed that in the case of the Blume-Capel model (for
which $\alpha_1=\alpha_2=0$), this SCOZA is different from the one 
presented in the main text. This can checked for instance by computing the
high-temperature expansion of the solution. Both theories yield
zero-field properties which are exact through order $\lambda^4$. It is
unclear which one provides the best numerical predictions.

\newpage

\section{}

In this appendix, we show that the tricritical scaling hypothesis, Eqs. (24) or
(26-27), is consistent with the asymptotic behavior of the SCOZA
PDE's, Eqs. (16). The notations are those of the main text.

To this end, it is convenient to rewrite the PDE's in terms of the two
variables  ${\cal E}=(1-z)^{1/2}$  and    $\rho=1-x$.  ${\cal E}^2$   is
proportional to the  inverse susceptiblity $\chi_0^{-1}$ and  $\rho=G({\bf
r}={\bf 0})+m^2=P(z)/c_0+m^2$.  At the TCP, we have  ${\cal E}= 0$ and
$\rho=\rho_t$. For $t \to 0$, $m\to 0$, and $\rho\to \rho_t$, Eqs. (16) take
the asymptotic form
\begin{mathletters}
\bea
\frac {P(1)^2} {\rho_t\lambda_t} \frac {\partial {\cal E}^2} {\partial t}
&=& 1 - (\frac {P(1)-1} {2}) [ \frac {b
\rho_t} {P(1)-1} \frac {\partial^2 {\cal E}} {\partial  m^2}-\frac {\partial^2 \rho} {\partial m^2}]\\
\frac {\partial \rho } {\partial t}& =& \frac {1} {2} {\lambda_t(1-\tau_t)}\frac
{P(1)-1} {P(1)}[\frac {b\rho_t} {P(1)-1}
\frac {\partial {\cal E}} {\partial  \delta \tau}-\frac {\partial \rho} {\partial \delta \tau}]
\eea
\end{mathletters}
where $t=(T-T_t)/T_t$,  $\lambda_t=cJ/(k_BT_t)$,  $\delta\tau=(\tau-\tau_t)/ \tau_t$,  and we have used
the expansion  of  the $3$-d   lattice  Green function   for $z\to  1$,
$P(z)\sim  P(1)[1-b{\cal E}+ O({\cal E}^2)]$, where $b$ is a positive
constant\cite{J1972}. In these  equations and  in the following,   all
derivatives are taken at the TCP.
 
By suitably rescaling the variables as ${\cal E} \to b \rho_t/ (P(1)-1)\
{\cal   E}$,  $t  \to   \lambda_tb^2\rho_t^3/[P(1)(P(1)-1)]^2    \ t$, $m   \to
m/(P(1)-1)^{1/2}$  and $\delta  \tau \to  b^2\rho_t^3/[P(1)(P(1)-1)^3(1-\tau_t)]\
\delta\tau$, we obtain the two simplified equations
\begin{mathletters}
\bea
\frac {\partial {\cal E}^2} {\partial t}&=&1 -\frac {1} {2} \frac {\partial^2({\cal E}-\rho)} {\partial m^2}
\\
\frac {\partial \rho } {\partial t}&=& \frac {1} {2} \frac {\partial(
{\cal E}-\rho)} {\partial \delta \tau} 
\eea
\end{mathletters}

We now introduce the tricritical scaling ansatz for ${\cal E}$ and for
the singular part of $\rho$, according to Eqs. (26) and (27),

\be
{\cal E}\approx |t|^{\gamma/2} E_{\pm}(u,v )
\ee

\be
\rho_{sing}\approx|t|^{2-\alpha-\phi} R_{\pm}(u,v)
\ee
where $u=\frac {g} {|t|^{\phi}}$ and  $v=\frac {m} {|t|^{\beta}}$.  Because
of thermodynamic self-consistency, $E_{\pm}$ and $R_{\pm}$ obey
\be
\frac{\partial E^2_{\pm}}{ \partial u}=-\frac{\partial^2 R_{\pm}}{ \partial v^2}  \ .
\ee 

For the sake of  simplicity, we  only consider  the case $t>0$  (i.e.,
$T>T_t$),  but  a similar analysis can  be  performed for  $t<0$. Then
Eqs. (B2) yield
\begin{mathletters}
\bea
t^{\gamma -1}  [\gamma E^2_+-\phi u  \frac {\partial E_+^2} {\partial  u} - \beta v \frac {\partial
E_+^2 }{\partial v} ]+a(\phi-1)t^{\gamma-\phi}  \frac {\partial E_+^2}  {\partial u} =\nonumber\\
1 -\frac {1}{2} [ t^{\gamma/2 -2\beta } \frac
{\partial^2E_+} {\partial v^2}-t^{\gamma-\phi} \frac {\partial^2 R_+} {\partial v^2}]\\
 t^{1-\alpha} [(2-\alpha-\phi)R_+  -\phi u\frac {\partial R_+}
{\partial u} -\beta  v\frac {\partial R_+} {\partial  v}] +a(\phi-1)t^{2-\alpha -  \phi }\frac {\partial R_+}{\partial u}=\nonumber\\
\frac {1}  {2}   [t^{\gamma/2}\frac {\partial E_+} {\partial u}-t^{2-\alpha -\phi}  \frac  {\partial  R_+} {\partial u}]
\eea
\end{mathletters}
where we have used the   scaling relation $\gamma=2-\alpha-2\beta$ which  results
from thermodynamic  self-consistency.  If the cross-over  exponent $\phi$
is greater  than $1$ (which is  usually  the case and is  indeed found
numerically), the first  term in the left-hand  side of Eq.  (B6a) may
be neglected asymptotically. A non-trivial scaling  is then found when
the  exponents are related through the  two relations $\gamma=\phi=4\beta$. For
the same reason we may neglect the first term in the left-hand side of
Eq.  (B6b) and we obtain  the relation $\gamma=2(2-\alpha -\phi)$ (which
is not independent from the preceding ones). Actually, we expect that
as in the SCOZA for the Ising model, the enthalpy is analytic in $m^2$
and $T-T_c$ when approaching a critical point from a disordered phase,
which corresponds to $\gamma=2$ and $\beta=1/2$.(At the tricritical point,
the $m^2$ term of course vanishes.) The scaling
functions satisfy the two non-trivial PDE's
 
\begin{mathletters}
\bea
a(\phi-1)\frac {\partial E^2_+} {\partial u} &=& 1-\frac {1} {2} \frac {\partial^2 (E_+-R_+)} {\partial v^2}\\
\frac {\partial E_+} {\partial u} &=& [1+2a(\phi-1)] \frac {\partial R_+} {\partial u}
\eea
\end{mathletters}
 Using Eq. (B5) to eliminate one of the functions, we finally obtain a
single equation for $E_+$ 
 
\be
\frac {\partial E_+^2} {\partial u} = 1 - \frac {1} {2} \frac {\partial^2 E_+} {\partial v^2}
\ee
where we have used the rescaling $u\to 2/[2a(\phi-1)+1]  \ u$. (Note that
the multiplying factor is positive since $a>0$ and $\phi>1$.) Of course, this
equation  must be accompanied  by some boundary conditions.  These are
obtained  from the analytical requirements  that the scaling functions
must satisfy near the  TCP and in the  vicinity of the critical  lines
and the coexistence surfaces (see, e.g., Ref.\cite{LS1984}). It can be
shown  that  these boundary conditions are    also compatible with the
SCOZA  equations.    The other  function    $R_+$ can be   obtained as
$[1+2a(\phi-1)]R_+(u,v)=E_+(u,v)-v^2+R_{+0}$,   where $R_{+0}$  is  a constant
that can be determined by the boundary conditions.

Eq.  (B8) has  an important consequence for  the scaling behavior near
the TCP when one approaches the $\lambda$-line along a  path  at fixed $\tau$.
The convenient  temperature variable is then $\dot{t}=[T-T_{\lambda}(\tau)]/T_t$
which  measures the     distance   from  $\lambda$-line at fixed $\tau$, so that 
$t=\dot{t}+t_{\lambda}$, where   $t_{\lambda}=(T_{\lambda}(\tau)-T_t)/T_t$   defines  the
$\lambda$-line near the   TCP. When  $|\dot{t}|\to 0$, or $t\to t_{\lambda}(\tau)$, the
scaling field $g$ behaves as $g- g_0 \sim \dot {t}$, where $g_0=(\tau-\tau_t)/ \tau_t$
is a constant, and the scaling variables $u$ and $v$ behave as
$u-u_0\sim \dot{t}$ and $v\sim m$, where $u_0$ is a constant. As a
consequence, Eq. (B8) can be rewritten as

\be
\frac {\partial E_+^2} {\partial\dot {t}  } = 1 - \frac {1} {2} \frac {\partial^2 E_+} {\partial m^2}\ .
\ee  
This     is  precisely the asymptotic     SCOZA     equation for   the
spin-$\frac{1}{2}$  Ising   model   which     has been   studied    in
Ref.\cite{HPS2000} (with $E_+$ playing  the role of the  variable $\phi$
in that reference). Therefore,  one expects that the critical behavior
above   and below  the  critical   temperature  $T_{\lambda}(\tau)$  will  be
identical to that of the SCOZA for  the Ising model (with for instance
$\dot {\beta}=7/20$ for the zero-field magnetization $m(\dot{t}))$.

\newpage

FIGURE CAPTIONS: \\

Fig. 1: SCOZA spinodal surface of the 3-d
Blume-Capel in the ($T-\tau-m$) space.\\

Fig. 2 : Detail of the spinodal surface near the
TCP.\\

Fig. 3: Zero-field $T_c(\tau)$ phase diagram. Second-order and
first-order parts of the phase boundary are shown as full and dashed
lines, respectively. For numerical reasons, calculations
have not been performed below  $k_BT/(Jc) \approx 0.18$ (see text).\\

Fig. 4: Zero-field $T_c(\Delta)$ phase diagram. Second-order and
first-order parts of the phase boundary are shown as full and dashed
lines, respectively. The inset describes the vicinity of the TCP. \\

Fig. 5: Zero-field $T_c(x)$ phase diagram. Second-order and
first-order parts of the phase boundary are shown as full and dashed
lines, respectively. The inset describes the vicinity of the TCP.\\

Fig. 6: Temperature dependence of the order
parameter (full lines) and spinodal  lines (dotted) in the vicinity of
the TCP on the first-order part of the phase boundary. \\

Fig. 7: Projections of the wing boundaries onto the
$\Delta-T$, $\Delta-h$ and $T-h$ planes. \\

Fig. 8:  Log-log plot of the zero-field ordering susceptibility
$k_BT\chi_0$ as a function of the reduced temperature $1-T_c/T$ and
corresponding effective exponent $\gamma_{eff}$. The different curves
correspond to $\tau=1.00, 0.79, 0.60, 0.40, 0.29, 0.25$ and $0.22$.\\

Fig. 9:  Log-log plot of the order parameter $m$ as a
function of the reduced temperature $1-T/T_c$ in the 
second-order region well above the crossover to tricritical behavior.\\

Fig. 10: Log-log plot of the order parameter $m$ as a
function of the reduced temperature $1-T/T_c$ in the
second-order and tricritical regions.\\

Fig. 11 : Behavior of the wing boundaries as the TCP is approached: log-log plots of the critical field $h_c$ and
magnetization $m_c$ as a function of the
reduced temperature $1-T_c/T_t$.

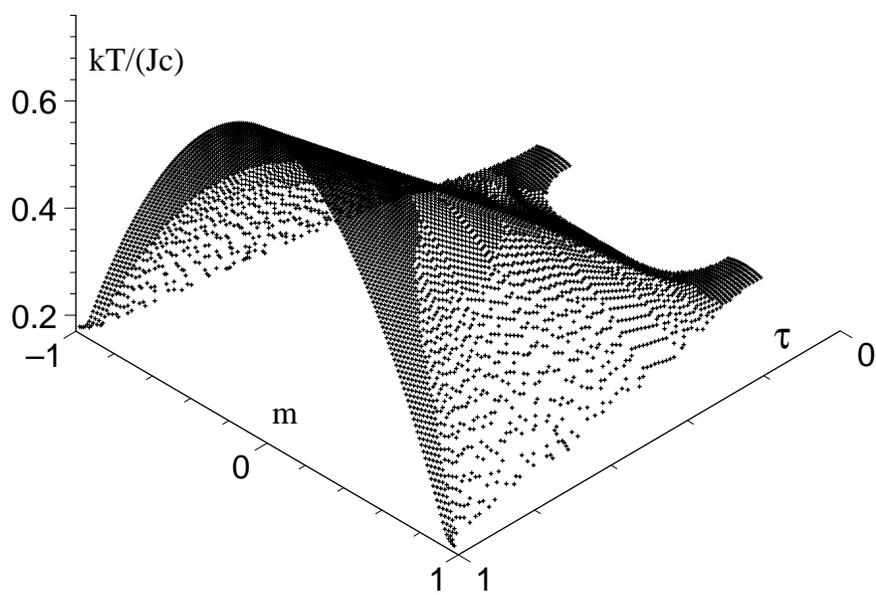
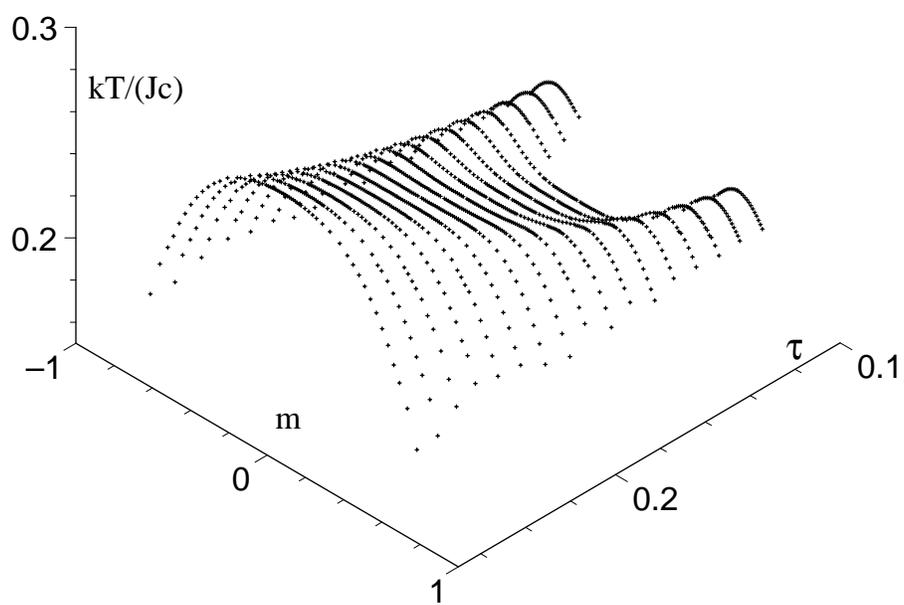
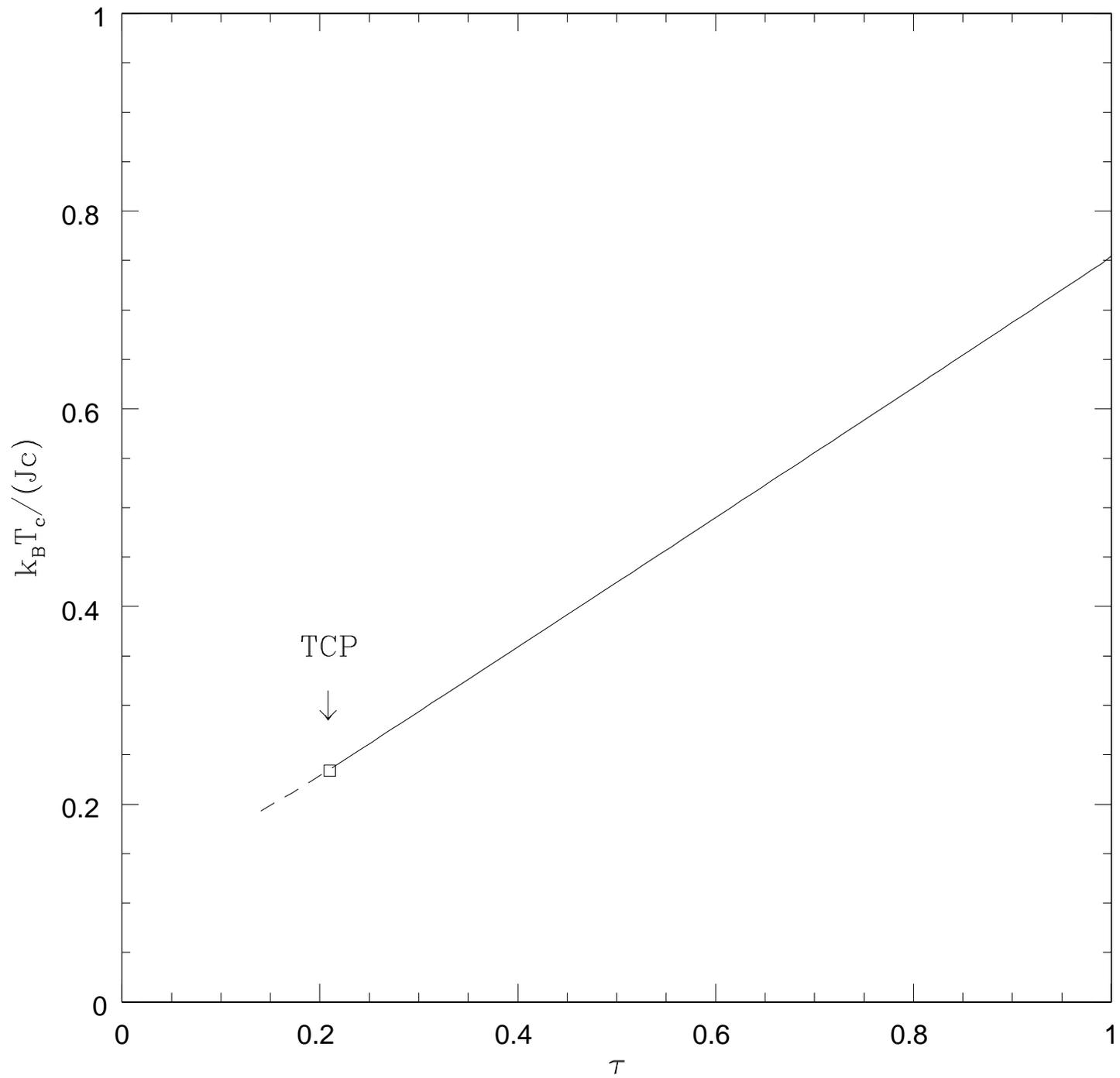
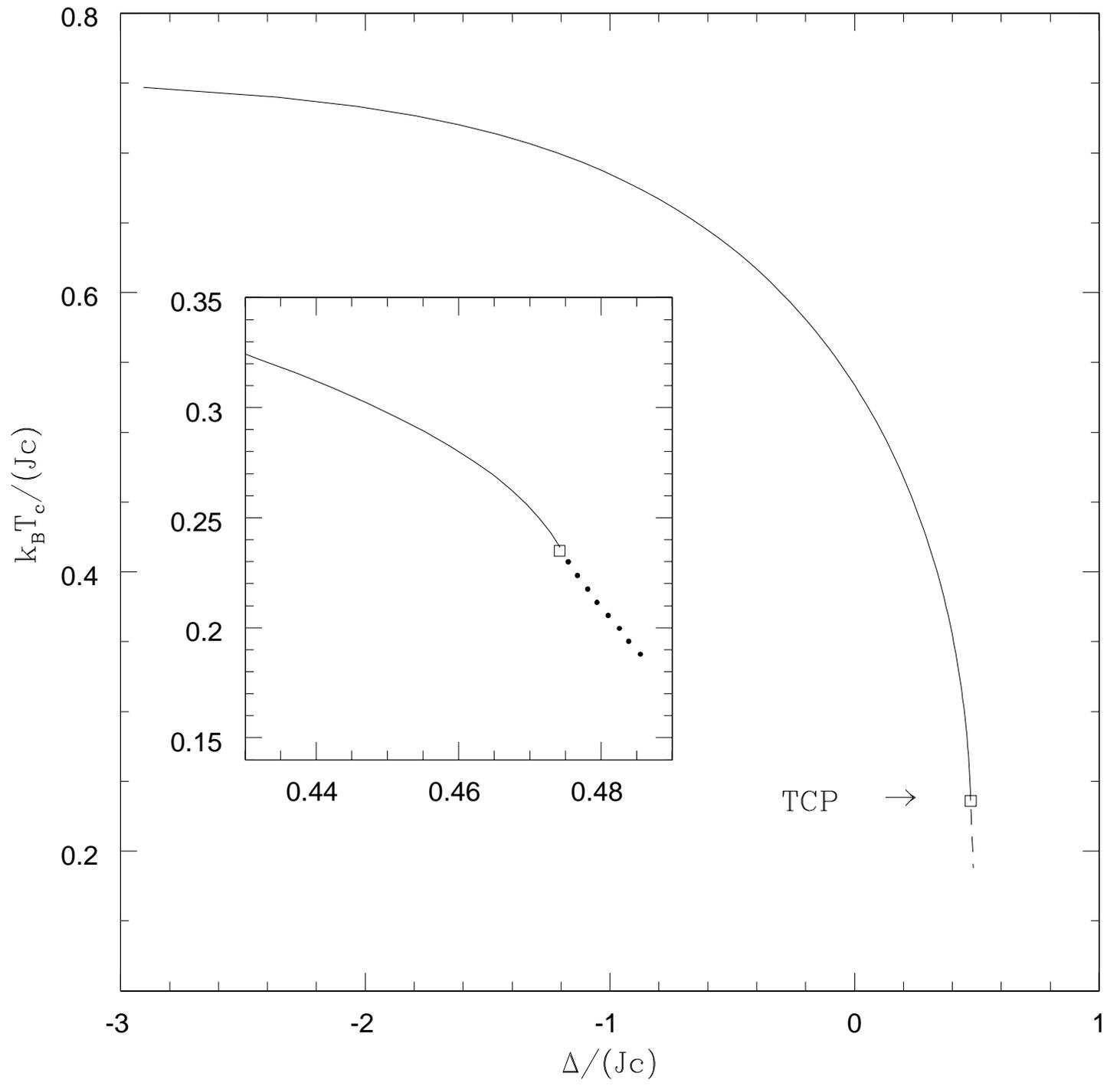
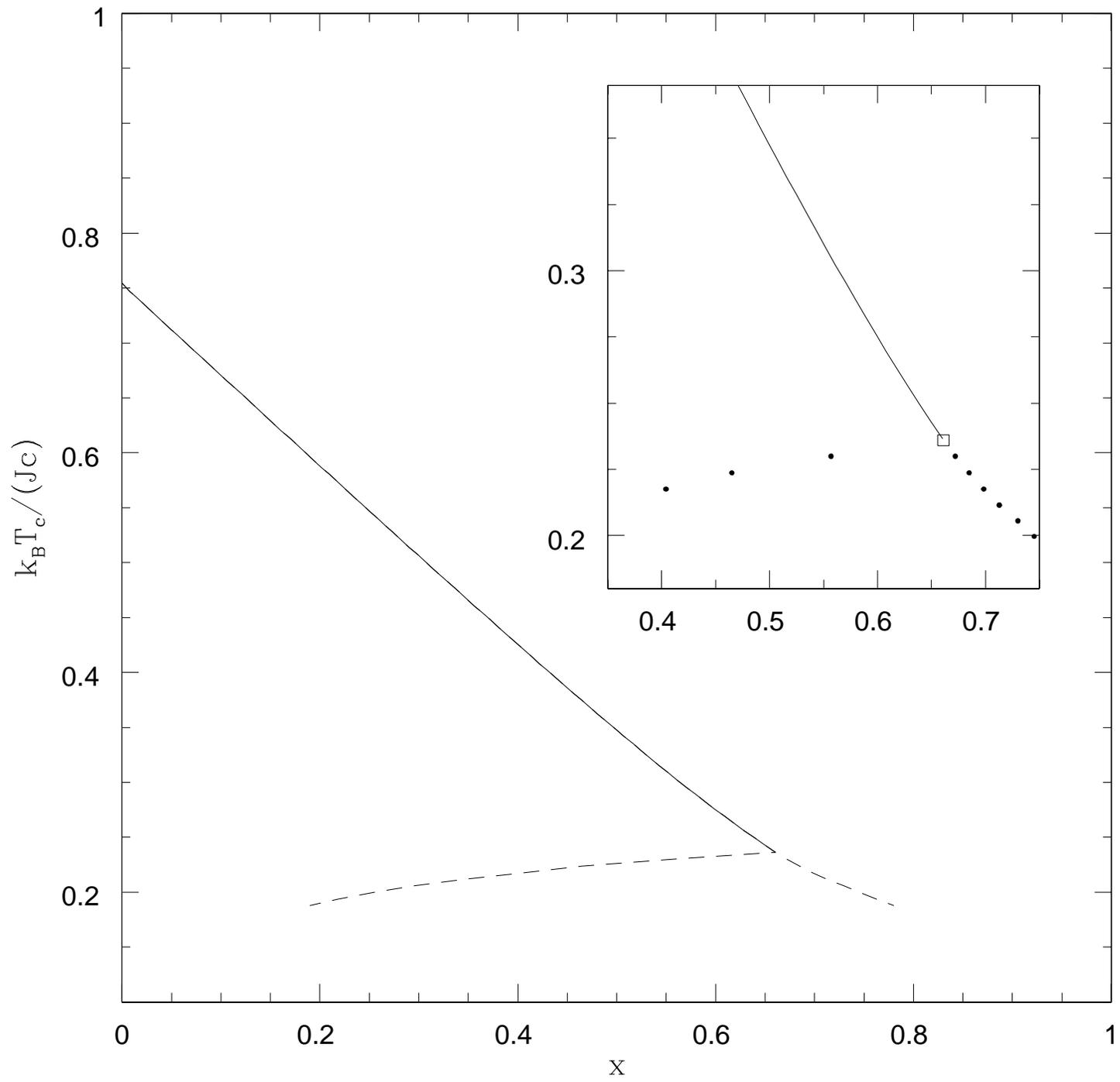
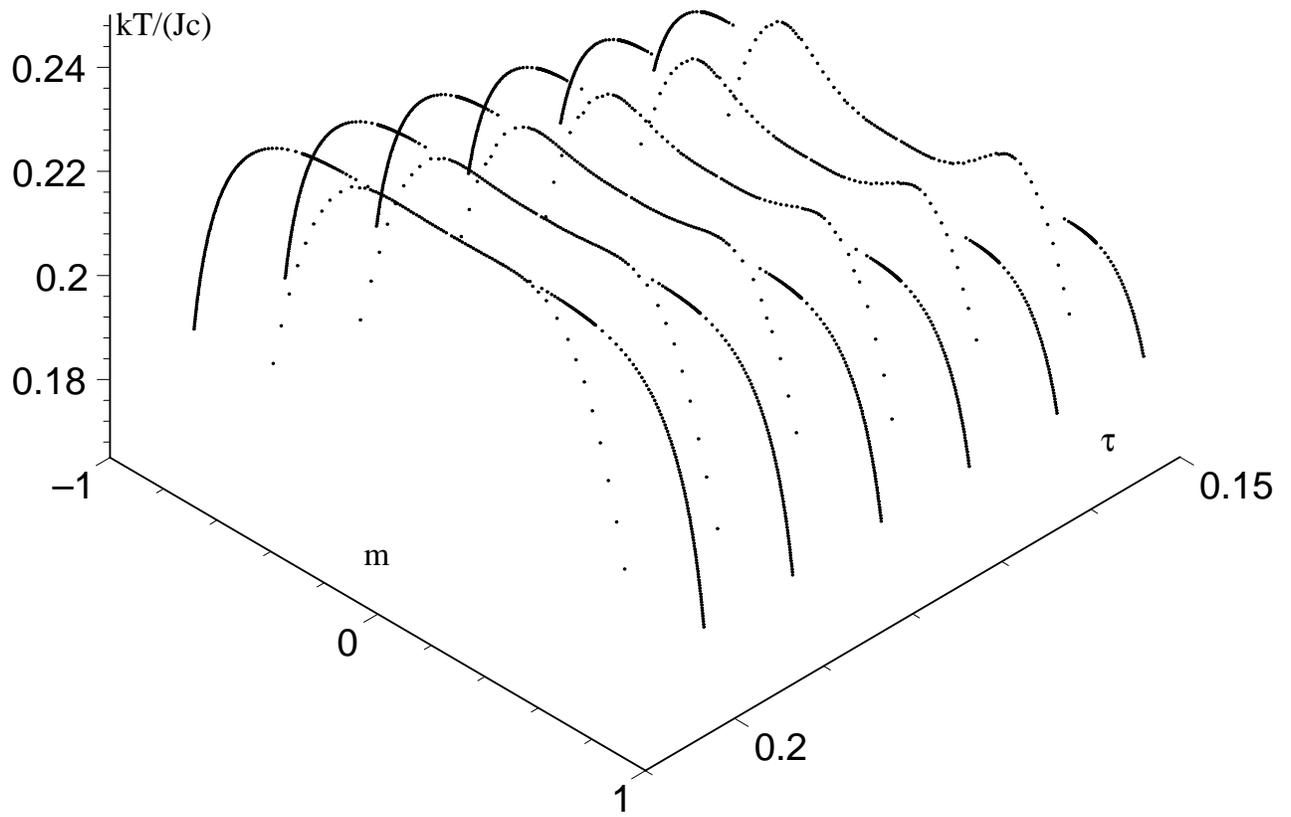
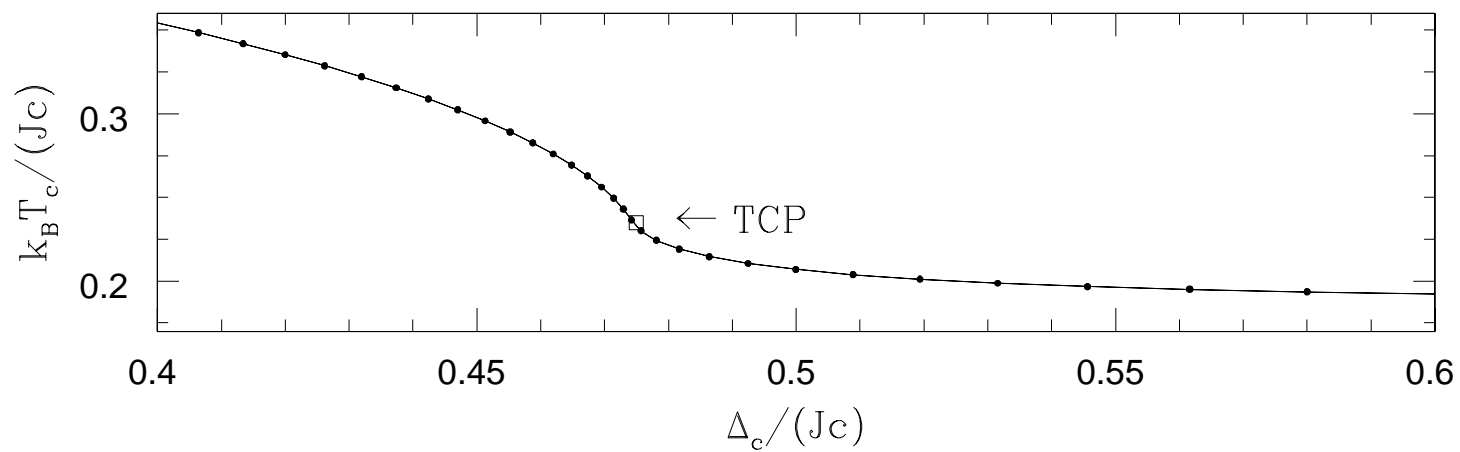
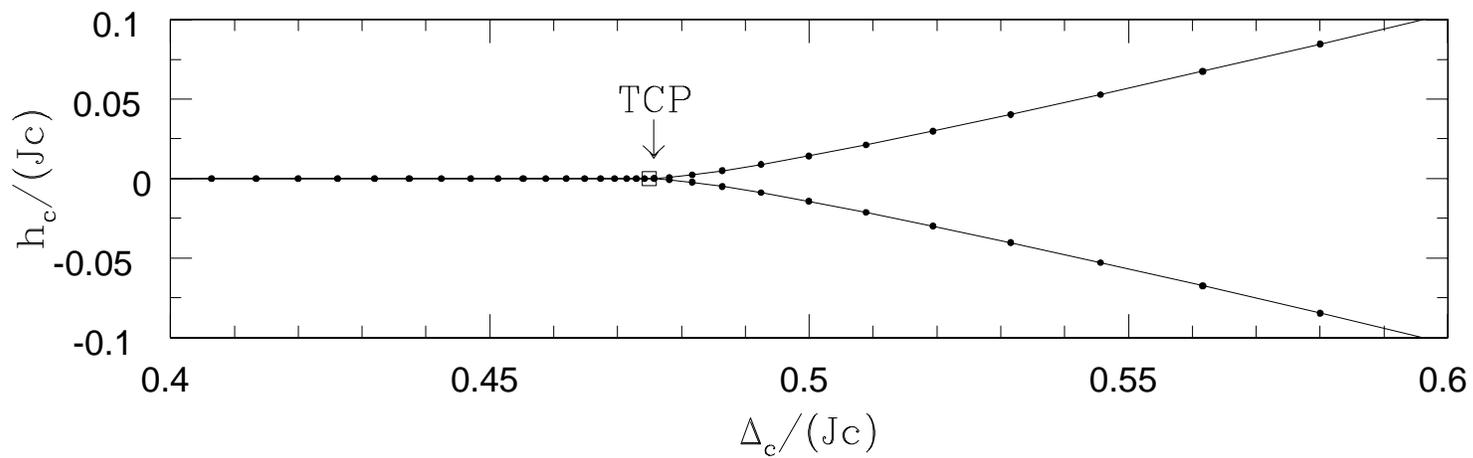
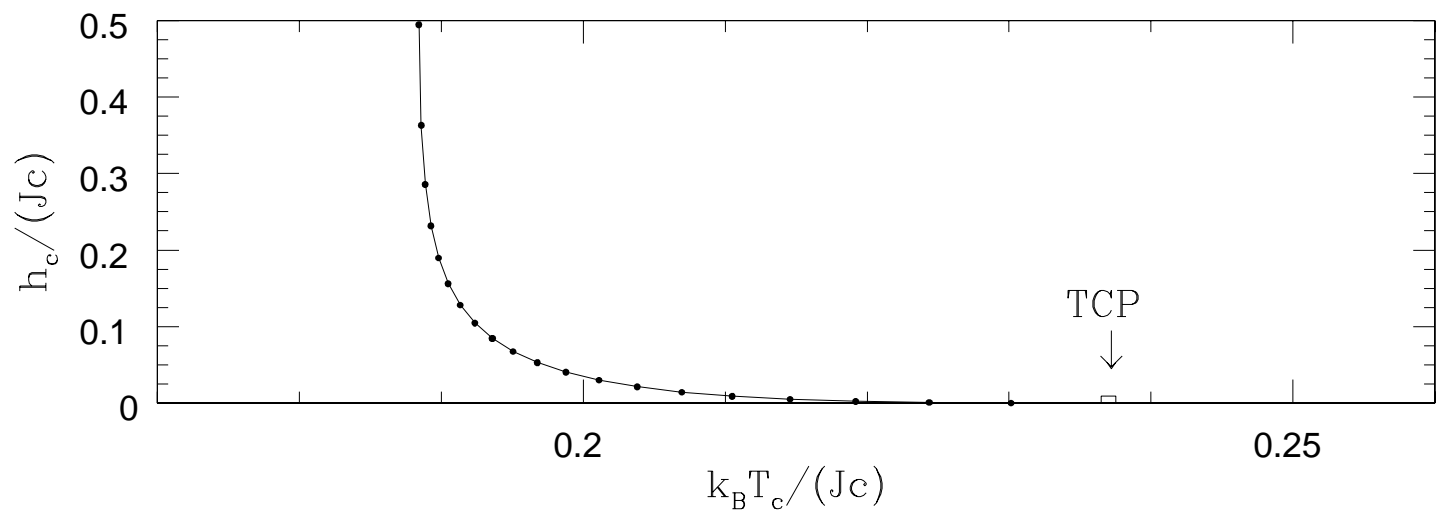
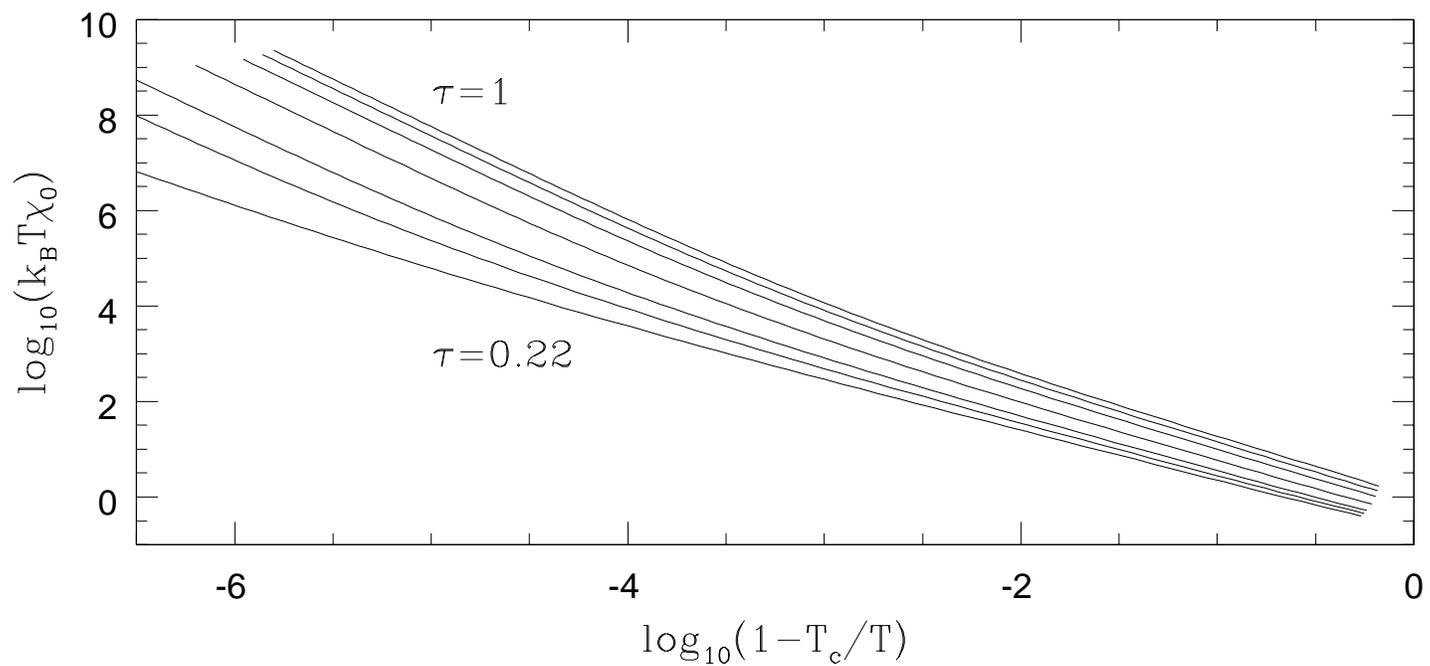
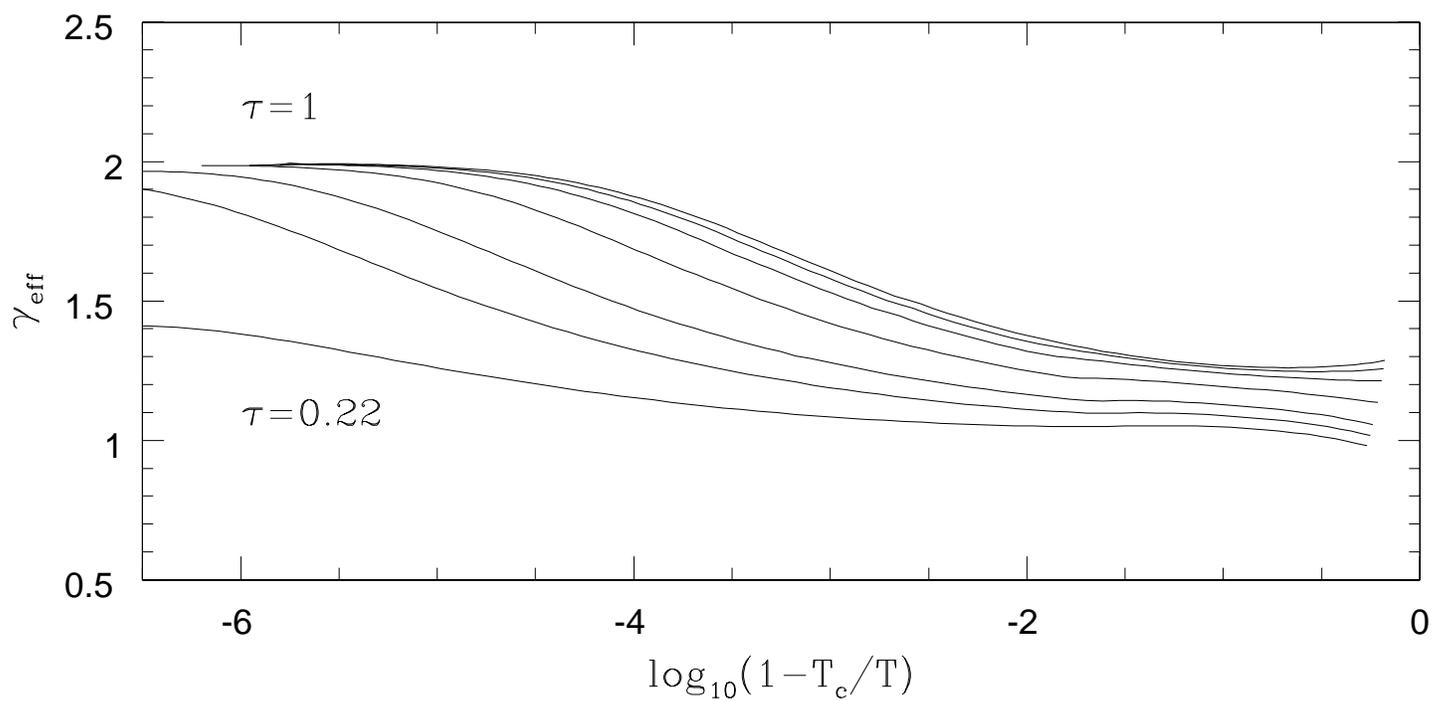
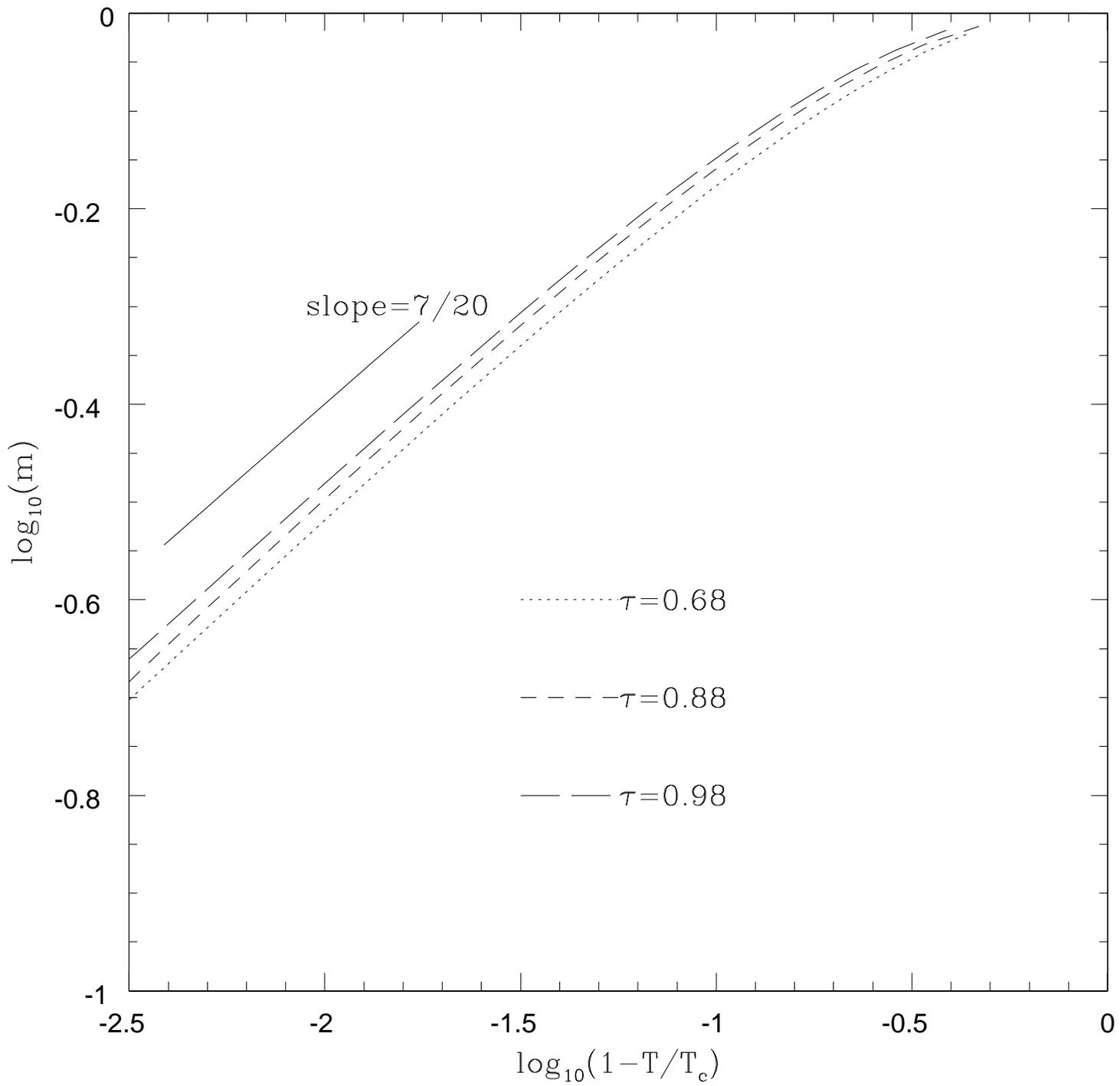
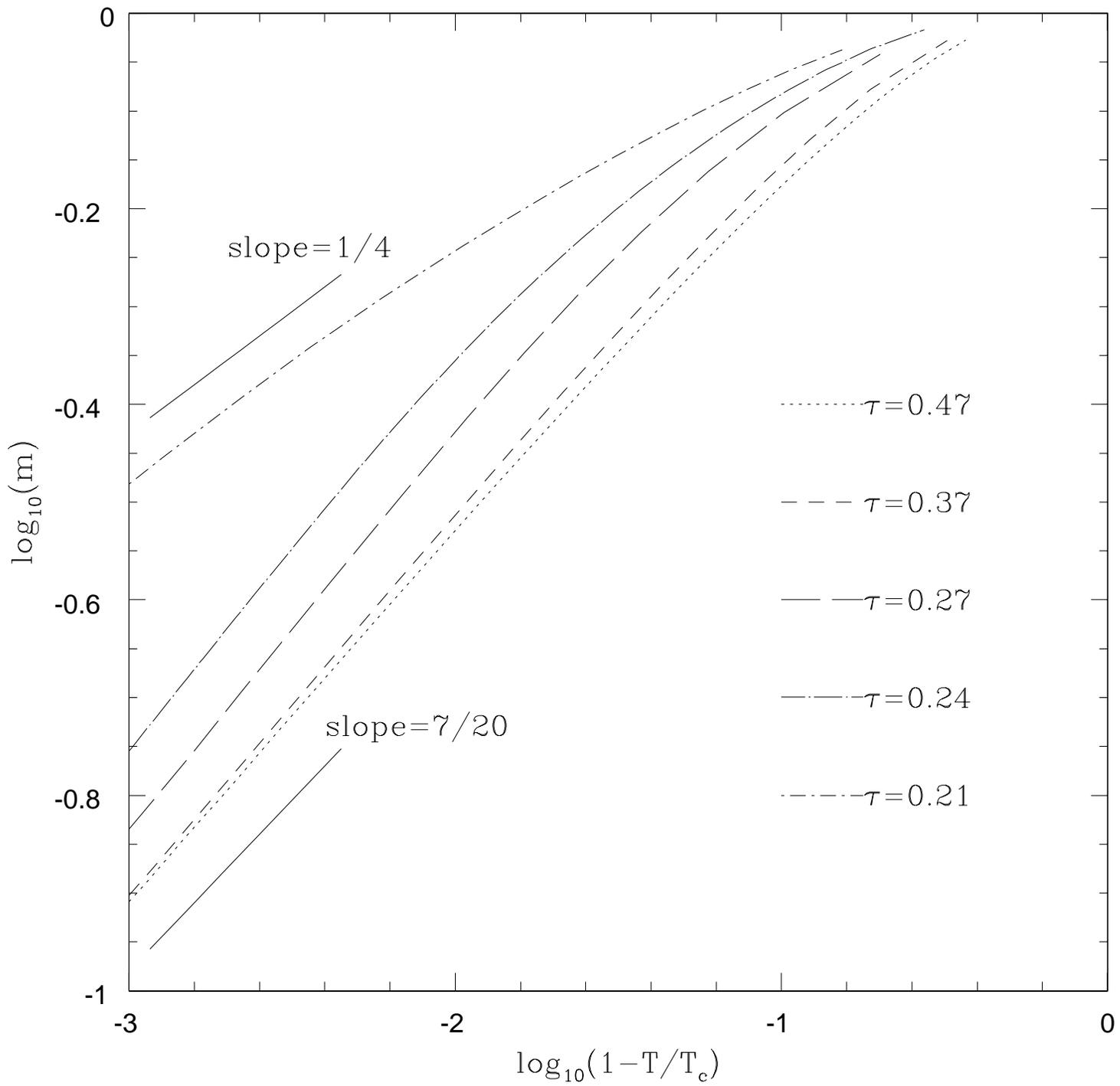
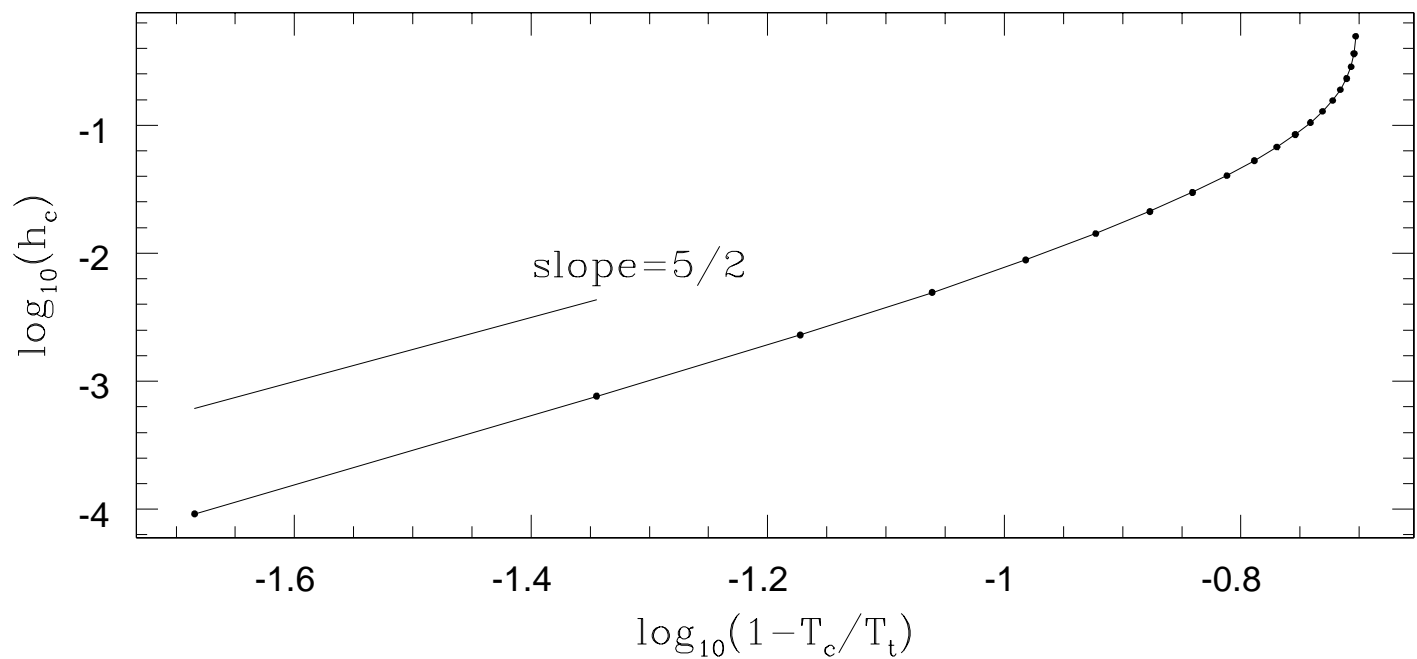
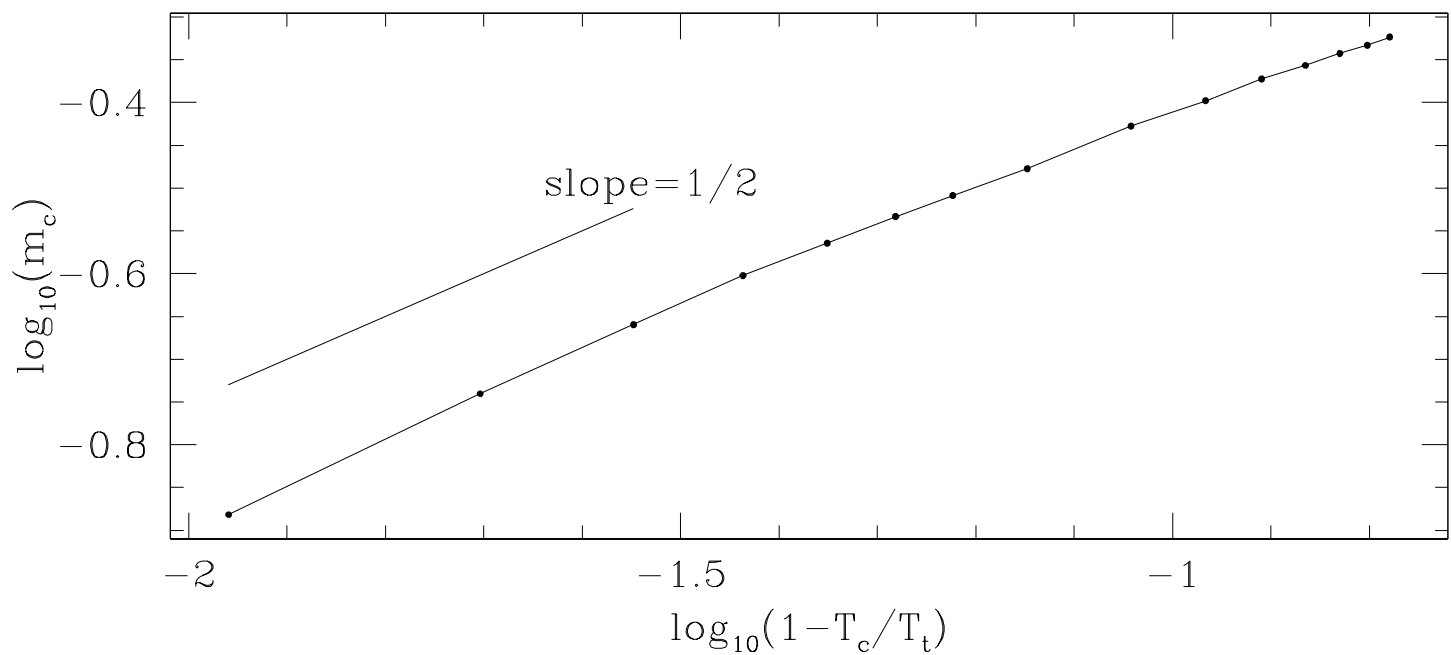
\end{document}